# Bribes to Miners: Evidence from Ethereum

Xiaotong Sun

University of Glasgow

## Abstract

In blockchain, bribery is an inevitable problem since users with various goals can bribe miners by transferring cryptoassets. To alleviate the negative effects of such collusion, Ethereum blockchain implemented new transaction fee mechanism in the London Fork, which was deployed on August 5$^{th}$, 2021. In this paper, we first filter potential bribery by scanning Ethereum transactions, and the potential bribers and bribees are centralized in a small group. Then we construct bribing proxies to measure the active level of bribery and then investigate the effects of bribery. Consequently, bribery can influence both Ethereum and other mainstream blockchains, in aspects of underlying cryptocurrency, transaction statistics, and network adoption. Moreover, the London Fork shows complicated effects on relationship between bribery and blockchain factors. Besides, bribery in Ethereum relates to stock markets, e.g., S&P 500 and Nasdaq, implying implicit interlinks between blockchain and traditional finance.

# 1. Introduction

Blockchain, as a *distributed ledger*, can execute transactions and update status without a trusted third party. This novel property, i.e., *decentralization*, is regarded as a disruption to centralized power. Benefiting from decentralization, many public blockchains have emerged and experienced rapid growth of adoption. In this paper, we focus on decentralization in a leading public blockchain, i.e., *Ethereum*.

In Ethereum, transactions are validated by *miners*. Currently, Ethereum applies *Proof-of-Work (PoW)* mechanism to generate consensus (Wood, 2021). In PoW blockchains, to add new blocks, miners need to solve complicated mathematical problems, i.e., PoW puzzles, and this process is usually called *mining*. As a result, miners with significant computational power are more likely to be winners of mining, and the powerful miners control over validation of most transactions. Then two questions naturally exist in PoW blockchains: is blockchain decentralized? What will happen if malicious users collude with dominant miners?

Centralization in blockchain is undeniable. Theoretically, Vitalik Buterin, the co-founder of Ethereum, proposes blockchain trilemma, claiming that decentralization, security and scalability cannot coexist in blockchain. Empirically, mining centralization exists (Gervais et al., 2014), and powerful miners can launch various attacks (Nakamoto, 2008; Teutsch, Jain, and Saxena, 2016; Eyal and Sirer, 2014). Furthermore, miners can exploit their power by receiving bribes (Bonneau, 2016), and bribery can be considered as collusion between bribers and miners. According to various goals of bribers, bribery can fall into several categories. Adversaries can bribe miners to revise the transaction history by forking the blockchain (Liao and Katz, 2017; Daian et al., 2020), and such attacks can widely affect users who have executed a series of transactions. Anti-blockchain bribers may even attempt to mine consecutive empty blocks (Bonneau, 2016) so they can devalue the blockchain. For bribers who focus more on their own interests, their aims include execution of certain transactions (McCorry, Hicks and Meiklejohn, 2018) and ignorance of others' transactions (Nadahalli, Khabbazian and Wattenhofer, 2021). Other possible bribery activities are discussed by Winzer, Herd and Faust (2019) and Judmayer et al. (2021a and 2021b).

Since collusion between miners and bribers can undermine health of blockchain, Ethereum proposed a new transaction fee mechanism in an improvement proposal, namely EIP-1559. The new mechanism aims to refrain collusion between miners and users (Roughgarden, 2020), and it was deployed to Ethereum on August $5^{th}$, 2021. Usually, the upgrade of Ethereum is called *the London Fork*, which is regarded as the most significant change to Ethereum so far. Leonardos et al. (2021) and Liu et al. (2022) discuss the influence of the London Fork, including fee, efficiency and security, however, no empirical analysis of collusion, especially bribery, has been presented. To fill the gap between bribery possibility and empirical evidence, this paper detects potential bribery in Ethereum, and the varying active level of bribery activities can help to examine practical effects of the London Fork.

To achieve that, we propose innovative methods to detect potential bribes to miners. In practice, it is difficult to detect if cryptoassets sent to miners are bribes. So, we filter all transactions that transfer *Ether (ETH)* to miners, since ETH is underlying cryptocurrency in Ethereum. To exact more information, we adopt one-step backward tracing to illustrate the circulation of ETH bribes. To summarize, by tracing the transaction history, we can extract transactions that include potential bribes to miners. Then, we establish proxies to measure the active level of bribery activities. The bribing proxies reflect on both motivations of bribers and efficiency of miners. For a briber, they may be more likely to collude with a miner when a certain goal is profitable enough. When a goal leads to more profits, a higher bribe will be worthy of attempt. If a briber's transaction is involved in a bribee's block, the signal of collusion is much stronger. It implies that the transferred ETH might be for execution of the certain transaction. On the other hand, the efficiency of bribable miners matters. In our bribing proxies, miners' efficiency is measured by the distance between bribery transaction and a new block validated by the bribee. A short distance means that the bribed miner can add a new block quickly after receiving bribes. Therefore, for potential bribers, a miner with centralized mining power will be an ideal choice.

We collect Ethereum data from January 1, 2019 to March 1, 2022. 982,116 transactions and 19,601 blocks are filtered, and 150 miners and 829 potential bribers are involved. The maximum of potential bribes is 7,620 ETH, and the maximum transferred value (in USD) is more than $12.5 million. Comparing to the rapidly growing blockchain users, the potential bribers and miners who receive transferred ETH are centralized in a small group, while the significant transferred value make these transactions more suspicious. Given the datasets for potential bribery, we construct bribing proxies, where varying active level of bribery activities is observed. In some blocks, the possibility of bribing is dramatically high, implying that bribery probably exists on these days. Moreover, by comparing bribing proxies before and after the London Fork, the active level of bribery dramatically decreases after the fork, implying that the new fee mechanism can refrain potentially malicious activities.

To investigate effects of bribery in blockchain, we choose some new categorized factors to capture the status of blockchain. Though some studies point several common factors of cryptocurrency (Liu and Tsyvinski, 2020; Liu, Tsyvinski and Wu, 2022), these factors are more related to risks and returns. So, our factor selection, e.g., network adoption and transaction statistics, can be reference for further research. The empirical framework brings forward some interesting findings. First, potential bribes can affect Ethereum and its underlying cryptocurrency, i.e., Ether (ETH). For example, higher active level of bribery can lower ETH price and market cap, but the transaction volume and the number of active users will be higher. Therefore, the influences of bribery can be complex. More interestingly, such complicated influences can be detected in other mainstream blockchains. For example, more bribery in Ethereum will lead to more transactions in Bitcoin, and more new users will be attracted. However, bribery in Ethereum can negatively affect Bitcoin price and market cap as well. These cross-chain effects point that blockchains should not be evaluated as independent markets, and beside short-term influences caused by price and volume, user flows among different blockchains may undermine the health of blockchains that are affected by vicious activities. Besides, potential bribes show interlinks with stock markets. Prices of S&P 500 and Nasdaq will decrease when potential bribes are more active, implying opaque but existing interlinks between blockchain and stocks. Previously, Liu and Tsyvinski (2020) find that the risk-return tradeoff of cryptocurrencies may not be correlated with stock markets. However, more implicit interactions may exist between cryptocurrency market and traditional financial market.

The remainder of our paper is organized as follows. Chapter 2 introduces background, including how Ethereum blockchain works and bribery in blockchain. In Chapter 3, we describe how to detect potential bribery and construct bribing proxies. The empirical results are presented in Chapter 4. Chapter 5 gives robustness checks, before Chapter 6 concludes.

## 2. Background

### 2.1 How mining in Ethereum works?

Ethereum is one of the most influential public blockchains, and the validation of transactions on Ethereum relies on a decentralized consensus mechanism, namely *Proof-of-Work (PoW)*. In PoW blockchain, *mining* is the process of adding a new block to the existing blockchain, and the participants are usually called *miners*. Mining resembles competition of computing power. To add new blocks, miners will compete in solving difficult cryptographic problems, i.e., PoW puzzles (Atzei, Bartoletti and Cimoli, 2017). Usually, miners with more computing power will be more likely to win the mining process, and only the winning miner will be able to add the next block and be rewarded. For each block, the rewards include a block reward and transaction fees paid by the transaction senders (Liao and Katz, 2017). Table 1 introduces more key variables and jargons in Ethereum.

PoW mechanism leads to intense competition, and a most widely noticed problem is mining concentration. Currently, the most blocks were added by a small group of miners (Gervais et al., 2014). To compete with powerful large miners, other individual miners can arrange themselves into "pools", i.e., *mining pools*. The integrated mining power contributes to a higher possibility of wining the mining process, and once a mining pool succeeds to add a new block, the rewards will be proportionally distributed among members. Yet, the emergence of mining pools further accelerates mining concentration (Gencer et al., 2018).

[Table 1 here]

## 2.2 Transactions on Ethereum

All actions on Ethereum are executed in the form of transactions. Figure 1 illustrates transaction execution on Ethereum. A transaction should be first broadcast to the *mempool*, which is like waiting area in blockchain. Then, miners will group waiting transactions in their blocks. Usually, the decision, including the transactions and their order, relies on the attached transaction fee (McCorry, Hicks and Meiklejohn, 2018). The chosen transactions will be executed once the block is appended to the existing blockchain. The decision power may result in involvement and exclusion of certain transactions, and miners could re-order transactions for their own profits (Daian et al., 2020).

[Figure 1 here]

## 2.3 Bribery in blockchain

As transaction validation is in charge of several most dominant miners, bribery, as a cheaper way to utilize mining power, logically exists in blockchain. Given different goals, bribery activities can fall into several categories. First, bribers can attempt to fork the current chain for their own interests. For example, in *history revision bribery*, bribers aim to rewrite blockchain history to steal a sizable sum of cryptocurrencies (Daian et al., 2020). Second, bribers are willing to pay bribes for transactions that can bring excessive gains (McCorry, Hicks and Meiklejohn, 2018). Third, bribers may attempt *ignore attacks* to exclude some transactions (Nadahalli, Khabbazian and Wattenhofer, 2021). Furthermore, the adversary can devalue the blockchain, e.g., mining consecutive empty blocks (Bonneau, 2016), by paying enough to miners. There are also other possibilities of bribery (Winzer, Herd and Faust, 2019; Judmayer et al., 2021a), and we refer readers to Judmayer et al. (2021b) for more details.

Beside attaching high transaction fees, bribery provides another way to get certain transactions validated. Given the limited block size, a natural question is: will be bribery lead to higher transaction fees? Theoretically, blockchain is a transaction-processing system, and Huberman, Leshno and Moallemi (2021) argue that miners cannot profitably affect fees. Kroll et al. (2013) claim that transaction fees are not even a crucial role in the long-run development of blockchain. On the other hand, Easley, O'Hara and Basu (2019) point out that transaction fees will be higher since users battle to get their own transactions included. Moreover, after the emergence of large mining pools, higher fees will be charged (Cong, He and Li, 2021). Though these papers do not address existing bribery, the demand of bribers will make users battle more intense. Therefore, we expect to see higher transaction fees when detecting more bribing activities.

From users' perspective, blockchain is a queuing system in which users submit transactions and wait for validation. Rational users will react to varying waiting time and transaction fees (Hassin, 1995; Kittsteiner and Moldovanu, 2005). If bribery exists in a queuing system with decentralized decision making, the system wants to get as many bribers as possible, and the higher fee will discourage non-bribers (Lui, 1985). If bribers get some priority, then both higher waiting costs and losses caused by bribery can negatively affect honest blockchain users. Even though bribers have different goals, the trust of blockchain community will decrease if users believe the health of blockchain is undermined (Böhme et al., 2015). So, we conjecture that bribery has negative effects on both network adoption of blockchain and transaction demand. Intuitively, users can reduce usage of the risky blockchain, and some of them may choose alternative blockchains. So, we expect to observe less transaction counts and lower transaction volumes when bribery is more active. Besides, if Ethereum users detect malicious bribery activities, they may choose other mainstream blockchains, e.g., Bitcoin.

Cryptocurrency markets can be probably affected by bribery activities. Recently, risk factors of cryptocurrencies attract much attention, and the discussion mainly focus on price- and market-related return predictors (Liu, Tsyvinski and Wu, 2022). In this paper, we investigate if certain users, i.e., bribers, can manipulate cryptocurrency performance, e.g., price, return, and volume. If such relationship exists, malicious user activities can expand characteristics used in investment models of cryptocurrencies. Furthermore, interlinks between cryptocurrencies and stock markets are also controversial. Liu and Tsyvinski (2020) argue that no direct relationship exists between cryptocurrency and stock markets. Since they apply standard textbook tools, on-chain activities, e.g., bribery, are not used to capture characteristics of cryptocurrencies. If bribery activities in blockchain show influence on stock markets, cryptocurrency markets are not isolated but a part of financial ecosystem.

To refrain problems caused by mining concertation and potential collusion related to transaction fees, Ethereum proposed the new transaction fee mechanism in a proposal named EIP-1559. Before EIP-1559, the fee mechanism is a first-price auction: transaction initiators propose a bid, and miners will get bids of executed transactions. In the new transaction fee mechanism, transaction fee will be divided into two parts, including base fee and a tip for miner. Once a transaction is validated, the base fee will be burnt, and the miner will only get the attached tip. Simply, since users can attach a tip, bribery should be less common under the new mechanism. More economic analysis is given by Roughgarden (2020). EIP-1559 was deployed on August 5$^{th}$, 2021, and the upgrade is usually called *the London Fork*. Empirical studies show that user experience is improved after the London Fork since waiting time decreases and estimating fee is easier (Liu et al., 2022). Given the positive effects of the London Fork, we expect that bribery is less active after the London Fork, though the effects of bribery are unpredictable.

## 3. Potential bribes to miners

### 3.1 Detection of potential bribes

Figure 2 illustrates the detection process of potential bribes to miners. Given a $block_i$ and a step length $step$, we examine some blocks prior to $block_i$. In these blocks, transactions are filtered if their recipient is $miner_i$. Senders of these transactions transfer an amount of value to a miner. Theoretically, any cryptocurrencies can be used in bribery (Judmayer et al., 2021b). In this paper, we only consider Ether (ETH), which is the underlying cryptocurrency of Ethereum. The attached value of ETH might be bribes, and the senders will be defined as potential bribers. In Figure 2, the bribers are $from_1, ..., from_n$.

Next, transactions in $block_i$ are checked. If potential bribers initiate a transaction in this block, the previously sent value is more likely to be bribes. Because the connected transactions imply that a briber first sends some value to $miner_i$, then the briber's transactions will be involved in $block_i$.

The pseduo-code (See Algorithm 1) has three input parameters, i.e., $startblock, endblock$, and $step$. The first two parameters set up the time interval of Ethereum dataset. $Step$ defines the number of scanned blocks prior to $block_i$. The output is datasets $transaction\_to\_block_i$, where $i \in (startblock, endblock)$, and the datasets include transactions detected as potential bribing.

[Figure 2 here]

[Algorithm 1 here]

### 3.2 One-step backward tracing

To better trace potential bribes, we apply on-step backward tracing algorithm. Figure 3 illustrates the general idea. Assuming that in $block_i$, $transaction_i$ includes potential bribes, blocks from $block_{i-1}$ to $block_{i-d}$ will be checked. In these blocks, transactions sent to the sender of $transaction_i$ are selected. More clearly, the pattern of connected transactions is $'miner_i \leftarrow address_A \leftarrow address'_B$, which illustrates the circulation of bribes in form of ETH.

Here, $d$, like the length of scanning window, is the number of checked previous blocks. By setting a short scanning window, the traced transactions are more likely related to potential bribing, i.e., transactions in dataset $transaction\_to\_block_i$. Finally, a dataset $trace_i$ will be generated for every $transaction\_to\_block_i$, where $i \in [startblock, endblock]$.

[Figure 3 here]

### 3.3. Proxies of active level of potential bribes to miners

In this section, we measure the active level of potential bribes to miners (referred to as 'bribing proxy') on Ethereum. We construct proxy benchmark based on transactions that include potential bribes, and proxies A and B are calculated by applying one-step backward tracing algorithm.

### 3.3.1 Proxy benchmark

Proxy benchmark measures the active level of potential bribes to miners (see Algorithm 2), using output of Algorithm 1. The numerator, i.e., $value$, reflects on the amount of bribes. Intuitively, higher values to miners are more likely to be bribes, and bribing attacks are related to transaction value (Judmayer et al., 2021; Somplinsky and Zohar, 2016). $Value$ also helps to exclude some legal activities. For example, a transaction will be automatically generated when a user joins a mining pool. This kind of transactions will not have an attached value, i.e., $value = 0$. Therefore, these transactions will not increase our proxy benchmark.

The denominator refers to the distance between potential bribes and the block validated by the bribee. With a longer distance, the correlation between value transfer and mining is weaker. In other words, the transferred value is less likely to be bribes. On the other hand, if the distance is short, the miner can be regarded as a 'efficient' bribee. Once the 'efficient' miner receives the bribes, bribers can expect their goal to be quickly achieved.

In the proxy, $weight_i$ can reveal a briber's real purpose to some extent. $Weight_i$ sums up the value of transactions initiated by potential bribers in $block_i$. When a briber executes a transaction in the bribee's block, the possibility of collusion should be different from the basic situation, i.e., no following transactions are involved in $block_i$. The value of transactions in $block_i$ reflects on the urgency of a user to execute a certain transaction. If a user is more urgent, he is more likely to bribe a miner.

Taken together, a block-level timeseries $p\_benchmark_i$ is established, where $i \in [startblock, endblock]$. Furthermore, a daily bribing proxy, i.e., $p\_benchmark_t$, can be calculated by summing up $p\_benchmark_i$ within a day $t$ (see Algorithm 3).

[Algorithm 2 here]

[Algorithm 3 here]

### 3.3.2 Validation of bribing proxy: one-step backward tracing

Proxy benchmark is improved by one-step backward tracing. In some cases, potential bribers tend to use a private smart contract to collude with miners, instead of directly sending bribes to mining pools (Judmayer et al., 2017). To validate the proxy, legitimate transactions should be excluded (See Algorithm 4).

Combining $trace_i$ with proxy benchmark, Algorithm 5 updates bribing proxies, and a new weight is introduced. The new weight can reflect on the relationship between $transaction\_to\_block_i$ and $trace_i$, and it is composed of two parts. One is 'distance', referring to the difference of block numbers between potential collusion and the connected earlier transaction. When two transactions are closer, the transactions are more likely to work for the same goal, and the goal of the group of transactions is more suspected.

The other part of new weight is about transferred value in connected transactions. When the value of a backward traced transaction is closer to potential bribes to the miner, the previous one is possible to be related to the potential bribing. By tracing prepositive transactions, we can partly conceal the real identity of the briber. The structure of our bribing proxies is given in Figure 4.

[Algorithm 4 here]

[Algorithm 5 here]

[Figure 4 here]

### 4. Empirical analysis of bribing proxies

This section summarizes the empirical results of this study. First, we describe the datasets for potential bribes on Ethereum and present descriptive statistics of bribing proxies. Then, we perform to investigate the effects of potential bribes in Ethereum. We consider both Ethereum and other three mainstream blockchains, including Bitcoin, Dogecoin, and Litecoin. For each blockchain, its underlying cryptocurrency, transaction statistics, and factors related to network adoption are studied. Besides, interlinks between potential bribes and stock markets are also considered. The description of factors is given in Appendix 1.

### 4.1 Data sources

On *Blockchair.com,* all on-chain transactions in Ethereum are publicly available. Cryptocurrency price, volume and market cap data are obtained from *Coingecko.com*, which aggregates financial data of most cryptocurrencies. Besides, *IntotheBlock.com* and *Etherscan.io* provide various statistics of mainstream blockchains, e.g., statistics of network adoption and transaction volumes. We extract Ethereum data from January 1, 2019 to March 1, 2021, including transactions from block 6988615 to block 14303536.

On August 5, 2021, *the London Fork* was deployed to Ethereum, meaning that transaction fee mechanism significantly changed (Roughgarden, 2020). The new transaction fee mechanism was proposed to alleviate collusion between miners and users. Currently, users can pay 'tips' to miners, therefore, users can get their transactions easily included by setting more 'tips'. The latest research focus on both theoretical models (Leonardos et al., 2021) and basic empirical analysis (Liu et al., 2022), but these findings ignore that users can bribe miners by simply transferring cryptocurrencies, which is hard to be refrained by mechanism design. So, in each theme of empirical analysis, we will examine the effects of new transaction fee mechanism.

### 4.2 Descriptive statistics of bribes to miners

In our analysis, $step$ is 1000, $d$ is 6000, and $c$ equals to 1. A small $step$ is taken in consideration of mining concentration. Currently, most blocks are added by a small group of miners, and most of them are mining pools. A smaller $step$ contributes to excluding some legitimate transactions, for example, a transaction sent to a mining pool when a user joins the pool. We select a relatively small $d$, meaning that the traced transactions will be more likely to relate to potential bribes.

From January 1, 2019 to March 1, 2022, 982,116 transactions and 19,601 blocks are filtered. The maximum of transferred Ether (ETH) is 7620, and the maximum transferred value (in USD) is approximately $12.5 million (See Table 2). Since most mining pool does not require a membership fee, these large transferred value is noteworthy and abnormal. After applying one-step backward tracing algorithm, we filter 11,352,816 transactions connected with potential bribes.

In potential bribing transactions, the participants are concentrated, including 150 miners and 829 potential bribers. Table 3 lists 20 most frequently involved miners. Beside leading mining pools, anonymous miners are also recognised as potential bribees. Table 4 highlights 20 potential bribers with the highest frequency, including mining pools, a smart contract of a crypto exchange, and anonymous users.

Then, we establish the proxies to measure the active level of potential bribes. Figure 5 shows that the active level of potential bribes is usually not very high, while spikes exist on some days, implying that suspicious activities may be implemented. By comparing descriptive statistics before and after the London Fork, we find that bribery activities are less active after the London Fork, which is preliminary evidence of effects of new fee mechanism.

[Table 2 - 5 here]

[Figure 5 here]

### 4.3 Underlying cryptocurrencies of blockchains

Ether (ETH) is the underlying cryptocurrency of Ethereum, so the financial factors of ETH may directly relate to bribes to miners. To capture the cryptocurrency performance, we consider price, daily return, volume (in native units) and market cap (in USD). If ETH is affected by bribery, transactions of other cryptocurrencies cannot be immune since ETH is the base payment in Ethereum. We also consider three underlying cryptocurrencies of other blockchains, including Bitcoin (BTC), Dogecoin (DOGE) and Litecoin (LTC). Theoretically, some bribing attacks will be implemented using several blockchains (Judmayer et al., 2021a), so potential bribes may have cross-chain influence. We estimate the following regressions:

$$token_{i,t} = \beta_0 + \beta_1 bribing_t + \beta_2 control_{i,t} + \beta_3 post_t \times bribing_t + \varepsilon_{i,t} \quad (10)$$

Where:

- $i = \{Ethereum, Bitcoin, Dogecoin, Litecoin\}$
- $bribing = \{benchmark, A, B\}$
- $token = \{Price, R, Vol, Mktc\}$
- $control = \{Active, BlockCnt, BlockTime, AvgFeeUsd\}$
- $post_t = \begin{cases} 0, t < Aug\ 5, 2021 \\ 1, t \geq Aug\ 5, 2021 \end{cases}$

In our regression models, we choose four control variables, including the number of active addresses, block count per day, the average time interval between blocks, and average transaction fee (in USD). The number of active addresses is a measurement of network adoption, contributing to cryptocurrency evaluation (Cong, Li and Wang, 2021; Sockin and Xiong, 2020). Since scalability of Ethereum might be influential on users' and miners' decision (Daian et al, 2020), we choose the number of blocks per day as a measurement. Average transaction fee (in USD) describes transaction costs. On the other hand, in traditional bribing attacks, bribers can collude with users by paying extremely high fees (Liao and Katz, 2017). As for the average time interval between blocks, it can reflect on waiting time of users. Theoretically, confirmation time of transactions can be related to bribing attacks (Judmayer et al., 2021a; Somplinsky and Zohar, 2016). But it is technically hard to get confirmation time for all on-chain transactions. Hence, we choose time interval between blocks to measure how frequent a dozen of transactions will be executed in Ethereum. Since both waiting time and transaction fee are publicly observable, these two measurements can influence users' decision. For example, given a certain blockchain, if the fee is too expensive, or waiting time is too long, rational users may discard the blockchain.

Table 6 brings forward some findings for the effects of bribery for ETH and BTC. Overall, all three bribing proxies can decrease ETH price and market cap, implying that valuation of underlying cryptocurrency in Ethereum will be negatively affected. A possible explanation is that some bribing attacks aim to devalue blockchains (McCorry, Hicks and Meiklejohn, 2018), so bribers will manipulate ETH price and market cap to weaken the reliability of blockchain. As a result, ETH holders may sell their crypto assets, and potential new users will be less likely to join the market. Since bribery seems to undermine health of Ethereum, we are curious if the London Fork can alleviate the negative effects of bribery. Surprisingly, though bribery exists under the new fee mechanism, the bribing proxies are positively related to ETH price and market price. It is hard to explain how bribery increases ETH price and market cap, and our conjecture is that bribers' aims may change after the new fee mechanism. In other words, by implementing unknown on-chain activities, bribers can benefit from higher ETH price or market cap. So, the London Fork is not uninfluential, though we need more time to observe its long-term effects. Besides, daily return of ETH will be higher when bribery is more active, and the effects still hold after the London Fork. This finding makes the bribers' goals opaquer.

For the case of BTC, we observe very similar results, showing that BTC price and market cap are affected by bribery on Ethereum. The findings are normal that blockchain users may initiate transactions on multiple blockchains, so Bitcoin and Ethereum, as two most widely adopted public blockchains, will be common choices for investors. The effects of the London Fork on BTC are also consistent with results for ETH, which can be further evidence of interlinks between Bitcoin and Ethereum.

[Table 7 here]

### 4.4 Transaction statistics of blockchains

All on-chain activities are implemented in the form of transactions, and transaction statistics are signals of adoption and growth of blockchains. So, we estimate the following regressions:

$$chain_{i,t} = \beta_0 + \beta_1 bribing_t + \beta_2 control_{i,t} + \beta_3 post_t \times bribing_t + \varepsilon_{i,t} \quad (11)$$

Where:

- $i = \{Ethereum, Bitcoin, Dogecoin, Litecoin\}$
- $bribing = \{benchmark, A, B\}$
- $chain = \{TxnVol, TxnVolUsd, TxnCnt\}$
- $control = \{Active, BlockCnt, BlockTime, AvgFeeUsd\}$
- $post_t = \begin{cases} 0, t < Aug\ 5, 2021 \\ 1, t \geq Aug\ 5, 2021 \end{cases}$

Here we consider three transaction-specific statistics, including transaction volume in native units, transaction volume in USD, and the number of transactions per day. These three transaction statistics can illustrate both scalability and prosperity of blockchain. Hypothetically, if bribery exists in Ethereum and has negative effects, users could choose other blockchains, and we will observe varying transaction statistics. To capture such cross-chain effect, we consider Bitcoin, Dogecoin, and Litecoin. On the other hand, transaction fees are proposed by users, though only miners can decide which transactions will be validated. So, if bribers can get their transactions executed more easily, transaction fees in blockchain may be affected as well.

Table 7 shows that transaction volume (in ETH) will increase when potential bribery is more active, but transaction count on Ethereum is not affected, implying that normal users do not discard Ethereum because of bribery. The finding is not surprising since bribing activities are hard to be detected (Nadahalli, Khabbazian and Wattenhofer, 2021). In other words, most blockchain users may not even realized the existence of bribery, unless they experience losses caused by bribery. So the transaction volume and transaction count will not be negatively influenced. Moreover, the London Fork does not affect positive relationship between bribery and transaction volume (in ETH).

For the case of Bitcoin, the results are different. First, bribing proxies show positive relationship with the number of Bitcoin transactions (See Table 8). It is to say, more bribery activities on Ethereum will lead to more transactions in Bitcoin. To some extent, Ethereum and Bitcoin are substitutes for each other. If users suspect they will suffer from bribery on Ethereum, they will choose to trade on their preferred blockchain. The results of Dogecoin are similar (in Online Appendix 2), which validates our conjecture of cross-chain influence of bribery. Moreover, the London Fork may not help to refrain the cross-chain effects. For example, we do not observe how the fork will weaken the relationship between bribery in Ethereum and Bitcoin transactions.

[Table 7 – 8 here]

### 4.5 Network adoption

Network adoption is crucial for blockchain, e.g., network effects can influence valuation of cryptocurrencies (Sockin and Xiong, 2020). In blockchain, everyone can freely have *addresses*, which resembles bank accounts in traditional finance. One can have as many addresses as they require, and no third party will require any files, e.g., identification. If one plan to leave a blockchain, they can simply sell cryptoassets in their addresses and stop transactions. Therefore, network factors of blockchain may be highly sensitive to status of blockchain. To investigate how bribery relates to network adoption, we estimate the following regressions:

$$Network_t = \beta_0 + \beta_1 bribing_t + \beta_2 control_{i,t} + \beta_3 post_t \times bribing_t + \varepsilon_{i,t} (12)$$

Where:

- $i = \{Ethereum, Bitcoin, Dogecoin, Litecoin\}$
- $bribing = \{benchmark, A, B\}$
- $network = \{Unique, New, Active, Active.Ratio\}$
- $control = \{Price, TxnVol, BlockCnt, BlockTime, AvgFeeUsd\}$

- $post_t = \begin{cases} 0, t < Aug\ 5, 2021 \\ 1, t \geq Aug\ 5, 2021 \end{cases}$

Here, we consider four network factors, including the number of unique addresses, new addresses, active addresses, and the proportion of active addresses to unique addresses. For each blockchain, we consider two new control variables, i.e., price of the underlying cryptocurrency and transaction volume (in native units). Intuitively, price and volume are signals of performance of blockchain, and users may react to different status of blockchain based on their beliefs and preference. For example, users may leave a blockchain when the price of its underlying cryptocurrency keep decreasing, or the transaction volume is extremely low.

For Ethereum, bribing proxies are positively related to active addresses and active ratio, and we give two possible explanations. First, since bribes are attached in transactions, the potential bribers will be counted as active addresses. On the other hand, other users may implement transactions to defend own profits. For example, as explain in Chapter 4.3, potential bribes are related to price and return of underlying cryptocurrencies. As a result, rational users will execute different strategies by initiating transactions, though they may not realize existence of bribers. After the London Fork, bribery activities show negative effects on active addresses in Ethereum, while the relationship between active ratio and bribery is not influenced. If more active addresses can be a good signal for blockchain adoption, bribery has more negative influence after the London Fork.

Then, we observe user 'flows' among difference blockchains. For the case of Bitcoin, we will observe more new users and more active users when bribery is more active in Ethereum. Dogecoin may benefit from bribery in Ethereum as well, since active ratio of Dogecoin will be higher with more bribery activites in Ethereum (See Online Appendix 3). It is normal again that rational users will tend to use other blockchains, assuming that potential bribes may cause losses of normal users. So, malicious activities in one blockchain may have positive influence, e.g., better network adoption, on other blockchains. But not all blockchains will benefit from such cross-chain effects in all dimensions. For example, when bribery in Ethereum is more common, the unique addresses of Dogecoin and Litecoin will decrease. Currently, cross-chain transactions are easier via latest technical tools, e.g., *Bridge* (Ethereum, 2022). It is to say, users can more easily transfer their crypto-assets to other blockchains, further enhancing substitutability of blockchains. So, users may have different evaluation of co-exising blockchains, and they will join or leave blockchains when Ethereum is not reliable enough.

[Table 9 – 10 here]

**4.6 Global stock markets**

The interlinks between blockchain and stock markets are not well investigated. Previously, Liu and Tsyvinski (2020) argue that risks and returns of cryptocurrency markets are independent on traditional financial market. However, in consideration of complexity of blockchain activities, we examine if suspicious activities in Ethereum can be related to stock markets. Here, we select four stock indices, including Standard and Poor's 500 (S&P 500), Nasdaq (NASDAQ), Nikkei 225 (N225), and The Shanghai Stock Exchange (SSE). We estimate the following regression model:

$$Stock_{i,t} = \beta_0 + \beta_1 bribing_t + \beta_2 control_t + \beta_3 post_t \times bribing_t + \varepsilon_{i,t} \quad (13)$$

Where:

- $i = \{S\&P500, NASDAQ, N225, SSE\}$
- $bribing = \{benchmark, A, B\}$
- $stock = \{Price, Vol, R\}$
- $control = \{Price, TxnVol, BlockCnt, BlockTime, AvgFeeUsd\}$
- $post_t = \begin{cases} 0, t < Aug\ 5, 2021 \\ 1, t \geq Aug\ 5, 2021 \end{cases}$

In regression (13), we use several Ethereum-specific factors as control variables. Ether (ETH) price and transaction volume are fundamental signals of blockchain performance. Block count per time and the average time between blocks can reflect on the scalability and efficiency of blockchain. Furthermore, the average time between blocks and average transaction fees can show the costs, i.e., waiting time and fee, faced by blockchain users. Intuitively, agents face a trade-off between stock markets and blockchain. If potential bribes undermine profits of non-bribers, these non-bribers may go back to stock markets, or at least execute certain transactions in stock markets.

Table 11 shows that more active bribery is related to lower price of S&P 500 and NASDAQ, while no significant results relationship exists in N225 and SSE. After the London Fork, such negative relationship does not disappear. Our findings imply that more implicit interlinks exist between Ethereum and stock markets, which are different from arguments by Liu and Tsyvinski (2020). However, it is hard to explain why bribery activities in Ethereum will lead to lower price of stock indices, implying opaque interactions between blockchain and traditional markets.

[Table 11 here]

**5. Robustness checks**

To bribe a miner, the value of transferred ETH is crucial. Intuitively, a low value of ETH is less likely to be a bribe. For that reason, in the datasets of potential bribes, transactions with low value are excluded, and the thresholds are 0.1 and 1. We construct proxies again, and the descriptive statistics are given in Table 12. Then, we estimate regression models with control variables in Chapter 4, and the results are presented in Online Appendix 4. After excluding low-value transactions, most results are consistent with our findings.

[Table 12 here]

**6. Conclusion**

Given the existing mining concentration, decentralization seems to be impossible in blockchain. Previously, both theoretical discussion (Abadi & Brunnermeier, 2018) and empirical evidence (Gervais et al., 2014) show that complete decentralization is illusion, and several dominant miners can account for validation of most transactions in Ethereum. In this paper, we further investigate another ignored way to exploit centralized mining power, i.e., bribery. After scanning Ethereum transaction history, we detect that the susceptive interactions between miners and bribers exist, and the circulation of bribes can be more precisely illustrated by tracing previous transactions connected with bribery. Since the participants of potential bribery are centralized in a small group, the significant amount of transferred cryptocurrency is even more suspicious, which might be signals of cooperation between miners and bribers.

To better illustrate the active level of bribery in Ethereum, we first construct bribing proxies and observe spikes, implying bribery is not common but may affect blockchain in certain blocks validated on certain days. Then, we investigate the influence of potential bribery in Ethereum, and categorized factor analysis is presented. First, ETH, as the underlying cryptocurrency of Ethereum, will have lower price and market cap when bribery is active, and such negative influence can be observed in BTC as well, which is a signal of cross-chain effects. However, in some aspects, other mainstream blockchains can benefit from bribery in Ethereum. For example, more bribery can lead to more Bitcoin transactions and active Bitcoin users. All these cross-chain effects satisfy theoretical arguments (Judmayer et al., 2021b), and we argue that blockchains can substitute for each other, especially when some malicious activities can undermine health of some blockchains.

Beside multiple blockchains, users can also invest in traditional finance. We investigate the interlinks between potential bribes and four stock indices, while the relationship is complex and opaque. For example, more bribery is negatively related to lower prices of S&P 500 and Nasdaq. Though Liu and Tsyvinski (2020) find that the returns of cryptocurrencies may not be correlated with stock markets, blockchain can interact with stock markets in a more implicit way. Besides, such interlinks may only exist in certain regions.

Our results should be interpreted with their limitations in mind. First, we do not consider transactions that transfer other tradable cryptocurrencies on Ethereum. Consequently, a proportion of bribing transaction is ignored. Although the detected potential bribes might be less, we do not involve other cryptocurrencies for the reason of precise valuation. The exchange rates of cryptocurrencies are rapid-varying, and the rates are not completely consistent on different Decentralized Exchanges (DEXes). Technically, it is almost impossible to assess the real-time value of cryptocurrencies.

Second, we (partly) ignore smart contracts specifically written for bribing. A dozen of papers (McCorry, Hicks and Meiklejohn, 2018; Judmayer et al., 2021a) propose smart contracts that help bribers to collude with miners more conveniently and fairly. It is to say, bribers may not directly transfer ETH to a miner but implement a bribe by creating a specific and anonymous smart contract. However, with the bursting growth of smart contracts, it is hard to analyse all of them and judge the real purpose of the issuers. So, in this paper, we may only capture the crucial part in bribery, i.e., the transaction where miners receive bribes.

Thirdly, the incentives of bribers and miners are not clear. Some miners may be involved in collusion without realizing the briber's real attention. Since verifying will consume computation power, miners will group transactions without verification (Luu et al., 2015). As for incentives of bribers, it is hard to measure their gains from a single transaction. Traditionally, bribers attempted to double spend their cryptocurrencies (Bonneau, 2016), but the intended impact of bribery, such as, transaction ordering, may be more complicated (Judmayer et al., 2021b). Furthermore, some of them will only pay bribes after some time (Nadahalli, Khabbazian and Wattenhofer, 2021), which makes it more difficult to understand their gains.

Finally, the long-term influence of potential bribes on Ethereum is still unclear. Though we investigate the effects of bribery after the London Fork, bribery may not be eliminated, and new problems may show up. So, how to refrain bribery, along with better mechanism design, is worthy of further discussion.

**Algorithms**

**Algorithm 1. Identify transactions for potential bribes to miners**

> **Algorithm 1 (Identify transactions for potential bribes to miners)**
>
> Input: $startblock, endblock, step$
>
> **For** $block_i$ in range ($startblock, endblock$):
>
>     **Filter** all $transaction_j$ in $block_{i-step}, \dots, block_{i-1}$ satisfying:
>
> $$\text{Recipient of } transaction_j = miner_i$$
>
>     **Return** a dataset $transaction\_to\_block_i$
>
>     **Filter all** $transaction_s$ in $block_i$ satisfying:
>
> $$\text{Sender of } transaction_s \text{ is in the senders of } transaction\_to\_block_i$$
>
>     **Return** a dataset $transaction\_in\_block_i$
>
> Output: dataset $transaction\_to\_block_i$ and $transaction\_in\_block_i$, where $i \in (startblock, endblock)$.

**Algorithm 2. Proxy benchmark**

---

**Algorithm 2 (Proxy benchmark)**

Input: dataset $transaction\_to\_block_i$ and $transaction\_in\_block_i$, where $i \in (startblock, endblock)$.

**If** $transaction\_in\_block_i$ is empty:

$$p\_benchmark_i = c \times \sum_t \frac{value_t}{|blockNumber_t - blockNumber_i|}$$

#$t$ refers to $transaction_t$ in $transaction\_to\_block_i$; $c$ is a constant.

**If** $transaction\_in\_block_i$ is not empty:

$$basis_i = c \times \sum_t \frac{value_t}{|blockNumber_t - blockNumber_i|}$$

#$t$ refers to $transaction_t$ in $transaction\_to\_block_i$. $c$ is a constant.

$$weight_i = (1 + \sum_s value_s)$$

#$s$ refers to $transaction_s$ in $transaction\_in\_block_i$.

$$p\_benchmark_i = basis_i \times weight_i$$

Output: a time series $p\_benchmark_i$, where $i \in [startblock, endblock]$.

**Algorithm 3. Establish a time-series for daily bribing proxy**

> **Algorithm 3 (Establish a time-series for daily bribing proxy)**
>
> Input: $p\_benchmark_i$, where $i \in [startblock, endblock]$
>
> **For** $block_i$ validated on date $t$
>
> $$p\_benchmark_t = \sum_i p\_benchmark_i$$
>
> #Here, $p\_collusion_t$ is collusion possibility on date $t$.
>
> Output: a time series $p\_benchmark_t$, where $t$ stands for date.

**Algorithm 4. Trace transactions prior to potential bribing**

> **Algorithm 4 (Trace transactions prior to potential bribing)**
>
> Input: dataset $transaction_{to_{block_i}}$, $d$
>
> **For** $transaction_j$ executed in $block_i$ in $transaction\_to\_block_i$:
>
> **Select** previous $transaction_s$ satisfying:
>
> $\quad transaction_s$ is in blocks from $block_{i-1}$ to $block_{i-d}$
>
> $\quad$ Recipient of $transaction_s$, i.e., $address_A$ is the sender of $transaction_j$
>
> **Return** a dataset $trace_i$, including all $transaction_s$
>
> Output: a dataset $trace_i$, where $i \in [startblock, endblock]$

**Algorithm 5. Update bribing proxy**

---

**Algorithm 5 (Update bribing proxy)**

Input: $txn\_to\_block_i, txn\_in\_block_i, trace_i$

**For** $i$ in range $(startblock, endblock)$:

    $block_i =$ block number of the fixed block

    **Select** $transaction_s$ in $txn\_to\_block_i$:

    **For** $transaction_s$:

    $block_s =$ block number of $transaction_s$

    $value_s =$ transferred value of $transaction_s$

    **Select** all $transaction_j, j = 1, 2, \ldots, n$ in $trace_i$ that is linked to $transaction_s$:

        $block_j =$ block number of $transaction_j$

        $value_j =$ transferred value of $transaction_j$

$$weight_{i,s,j} = \left(1 + \frac{1}{block_s - block_j} \times \frac{value_s}{|value_s - value_j| + \varepsilon}\right)$$

$$p\_bribing_{i,s,j} = p\_benchmark_{i,s,j} \times weight_{i,s,j}$$

\#Here, we take $\varepsilon = 10^{-18}$.

    $p\_bribing_{i,s} = \sum_j p\_bribing_{i,s,j}$

$p\_bribing_i = \sum_s p\_bribing_{i,s}$

\#Here, $p\_bribing_i$ is possibility of bribery in $block_i$.

Output: a timeseries $p\_bribing_i$, where $i \in [startblock, endblock]$.

# Figures

## Figure 1. Ethereum blockchain

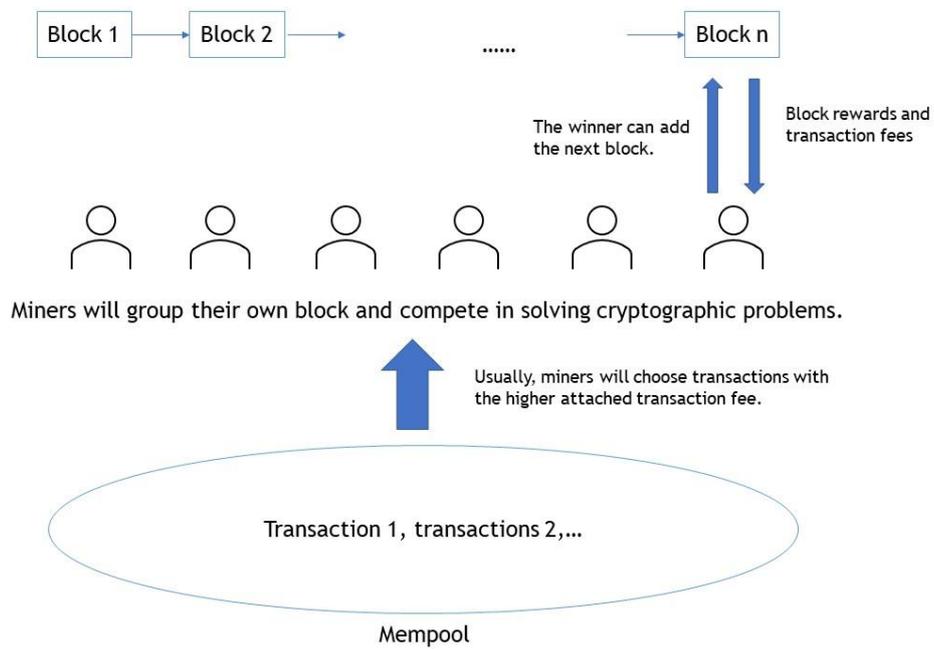

## Figure 2. Detection of potential collusion in Ethereum

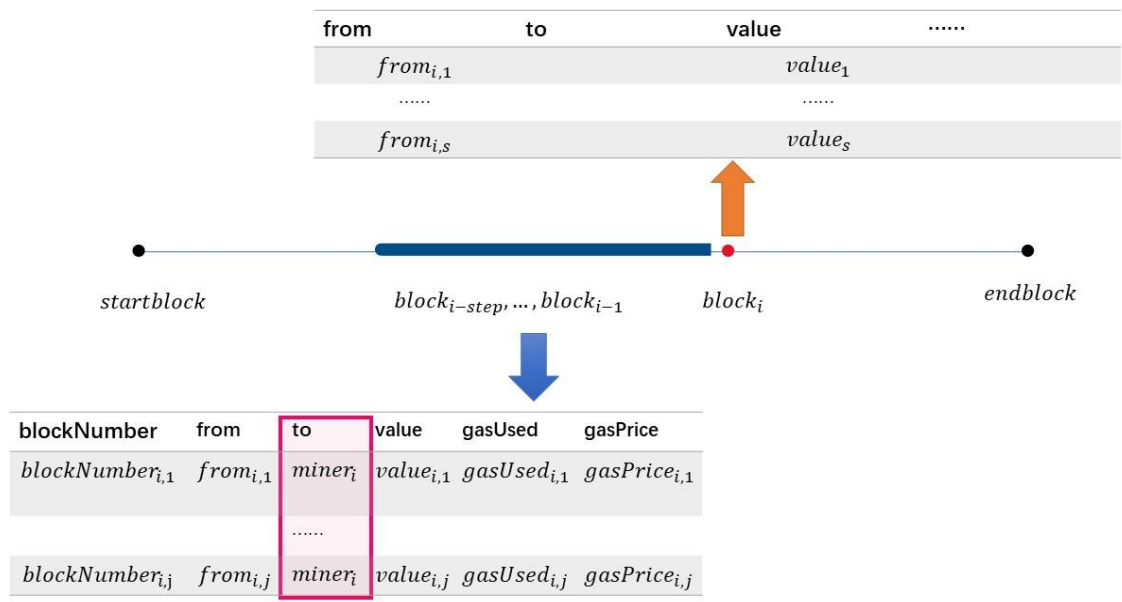

Note: This figure illustrates the detection of potential collusion. For a given block, e.g., *block i*, we will examine some previous blocks. Transactions sent to *miner i*, i.e., the miner of *block i*, will be filtered. The senders of these transactions will be regarded as potential bribers. Then, we will check transactions in *block i*. If potential bribers initiate transactions in *block i*, these transactions will be detected as a part of collusion as well.

**Figure 3. The idea of backward transaction tracing**

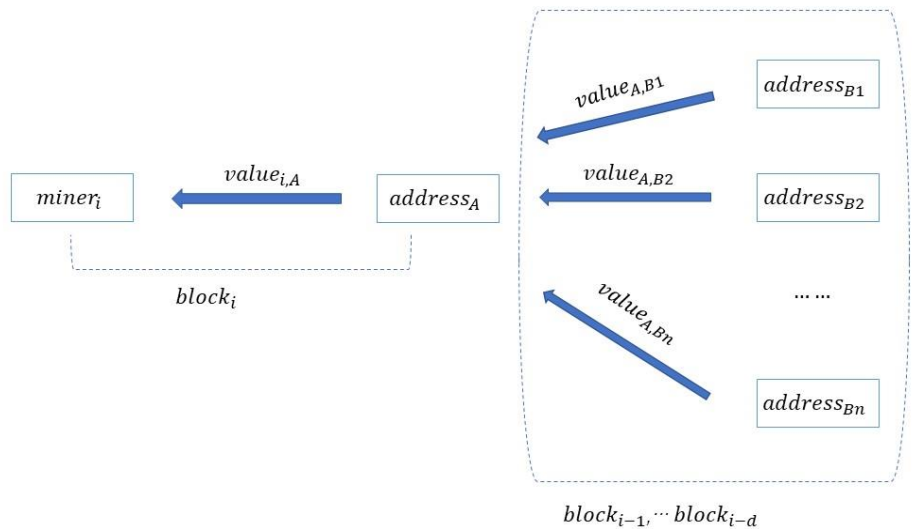

Note: This figure shows the one-step backward tracing algorithm. Given a transaction detected as potential collusion, i.e., a transaction in *block i*, we will check blocks $block_{i-1}$ to $block_{i-d}$. Assuming that address A is the potential briber, we will filter transactions sent to address A in the corresponding previous blocks.

**Figure 4. The structure of proxies of collusion possibility**

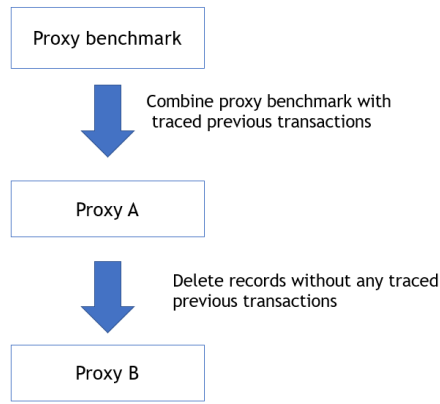

Note: This figure shows the structure of collusion proxies. We have proxy benchmark, proxy A and proxy B. Proxy A is developed after applying one-step backward tracing algorithm, and it could reveal more information, comparing to proxy benchmark. To calculate proxy B, we delete records without any traced previous transactions.

**Figure 5. Proxies of potential bribes (January 1, 2019 – March 1, 2022)**

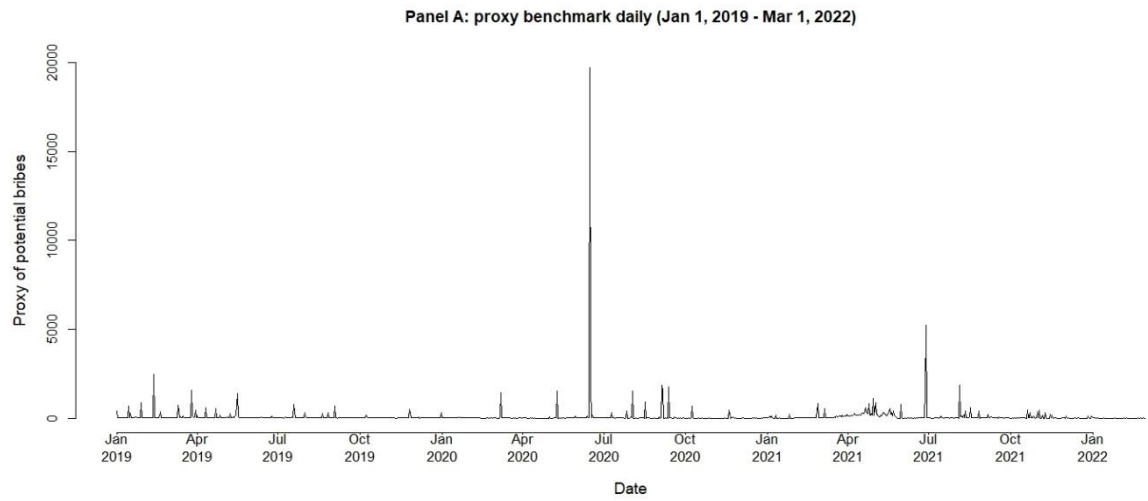

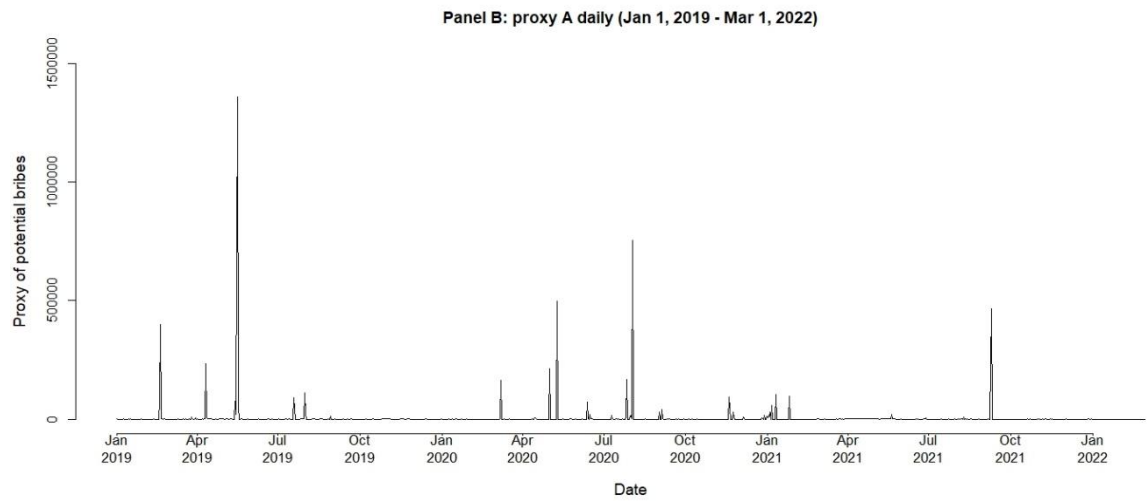

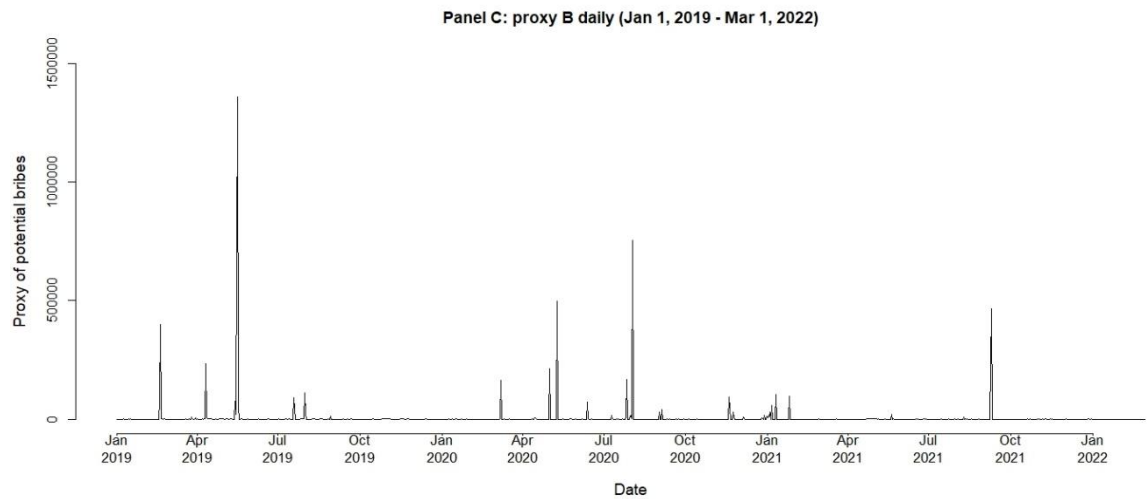

Note: This figure illustrates bribing proxies daily (January 1, 2019 – March 1, 2022). In Panel A, we present proxy benchmark, while proxy A and B are given in Panel B and C, respectively. Usually, the active level of potential bribes is not very high, while spikes exist on certain dates.

**Tables**

**Table 1. Terms of blockchain**

| Terms | Introduction |
| --- | --- |
| **Block** | Like a part of ledger, a block records some transactions pertaining to the blockchain. |
| **blockNumber** | The ordinal number of a block. For $block_i$, its block number is $i$. |
| **Miner** | The participant of mining process. The first one to solve a PoW puzzle can successfully add a new block. $miner_i$, refers to the validator of $block_i$. |
| **Startblock** | The start block in our data sample. |
| **Endblock** | The end block in our data sample. |
| **Address** | Accounts controlled by entities in Ethereum. Each account has a fixed address as the identity of the Ethereum account. |
| **Public name** | The name of an address. If an address has a public name, it is usually a smart contract of a DeFi or mining pool. |
| **Transaction** | A message with ETH and data from one account to another. |
| **From** | The sender's address of a transaction. |
| **To** | The recipient's address of a transaction. |
| **Value** | Transferred ETH of a transaction. |
| **Gas** | The computational cost of executing a transaction in Ethereum. |
| **GasUsed** | The units of gas actually used in a transaction. |
| **GasPrice** | The amount of ETH the sender is willing to pay per unit of gas. GasPrice is specified by the sender. |

**Table 2. Descriptive statistics of filtered transactions (January 1, 2019 – March 1, 2022)**

|  | Value (in ETH) | Value (in USD) | Fee (in Wei) | Fee (in USD) | Value_trace (in ETH) | Value_trace (in USD) |
| --- | --- | --- | --- | --- | --- | --- |
| **Mean** | 11.22 | 13718.49 | 5.29e+14 | 0.96 | 26.80 | 47448.77 |
| **Median** | 1.56 | 641.61 | 8.40e+13 | 0.05 | 1.00 | 1029.13 |
| **Max** | 7620.00 | 12550697.95 | 3.18e+16 | 103.85 | 249999.95 | 324298080.84 |
| **Min** | 0.00 | 0.00 | 0 | 0.00 | 0.00 | 0.00 |
| **Std** | 136.97 | 166421.29 | 1.19e+15 | 3.02 | 926.39 | 1843959.46 |

Note: This table reports the descriptive statistics of filtered transactions (January 1, 2019 – March 1, 2021). For the transactions detected as potential bribery, we investigate value (in ETH and USD) and transaction fee (in Wei and USD). $1\ ETH\ =\ 10^{18}\ Wei$. For the traced transactions that are connected with potential bribery, we consider value (in ETH and USD) of transactions.

**Table 3. The most frequently involved miners (January 1, 2019 – March 1, 2022)**

| Address | Public names | Freq | Mining Pool |
| --- | --- | --- | --- |
| 0xea674fdde714fd979de3edf0f56aa9716b898ec8 | Ethermine | 421245 | 1 |
| 0xb2930b35844a230f00e51431acae96fe543a0347 | MiningPoolHub: Old Address | 236888 | 1 |
| 0x3ecef08d0e2dad803847e052249bb4f8bff2d5bb | MiningPoolHub | 126376 | 1 |
| 0xd224ca0c819e8e97ba0136b3b95ceff503b79f53 | UUPool | 96083 | 1 |
| 0x5a0b54d5dc17e0aadc383d2db43b0a0d3e029c4c | Spark Pool | 28846 | 1 |
| 0xda466bf1ce3c69dbef918817305cf989a6353423 | MiningPoolHub: Old Address 7 | 28747 | 1 |
| 0x52bc44d5378309ee2abf1539bf71de1b7d7be3b5 | Nanopool | 12470 | 1 |
| 0x829bd824b016326a401d083b33d092293333a830 | F2Pool | 9439 | 1 |
| 0xf20b33875297687875451818387360290236704 | F2Pool | 5415 | 1 |
| 0x1ad91ee08f21be3de0ba2ba6918e714da6b45836 | Hiveon Pool | 2556 | 1 |
| 0x005e288d713a5fb3d7c9cf1b43810a98688c7223 | xnpool | 2516 | 1 |
| 0x04668ec2f57cc15c381b461b9fedab5d451c8f7f | zhizhu.top | 2096 | 1 |
| 0x00192fb10df37c9fb26829eb2cc623cd1bf599e8 | 2Miners: PPLNS | 1662 | 1 |
| 0x8595dd9e0438640b5e1254f9df579ac12a86865f | EzilPool 2 | 1325 | 1 |
| 0x9d6d492bd500da5b33cf95a5d610a73360fcaaa0 | Huobi Mining Pool | 862 | 1 |
| 0x35f61dfb08ada13eba64bf156b80df3d5b3a738d | firepool | 597 | 1 |
| 0xeea5b82b61424df8020f5fedd81767f2d0d25bfb | BTC.com Pool | 521 | 1 |
| 0x99c85bb64564d9ef9a99621301f22c9993cb89e3 | BeePool | 502 | 1 |
| 0x4c549990a7ef3fea8784406c1eecc98bf4211fa5 | Hiveon Pool | 354 | 1 |
| 0x0708f87a089a91c65d48721aa941084648562287 | Miner: 0x070...287 | 293 | 0 |

Note: This table reports 20 miners with highest frequency in the filtered transactions. If the miner has a public name, it is usually an address that belongs to some mining pool. Public names are accessed on *Etherscan.io*. In column 'Mining Pool', if an address belongs to a mining pool, the value will be 1. Otherwise, the value will be 0.

**Table 4. The most frequently involved senders (January 1, 2019 – March 1, 2022)**

| Address | Public Name | Freq | Entity |
|---|---|---|---|
| 0xf6da21e95d74767009accb145b96897ac3630bad | Ethermine: MEV Sender | 322620 | Mining Pool |
| 0xc168062c9c958e01914c7e3885537541dbb9ed08 | | 98288 | |
| 0x7d92ad7e1b6ae22c6a43283af3856028cd3d856a | UUPool: MEV | 95863 | Mining Pool |
| 0xafadc4302f07e9460eb4c31ec741c0f3e308ff3a | | 90530 | |
| 0xbfea450a21484539de16c1371a63a8bd681dc5bf | | 77740 | |
| 0xb0a3998133940095351f32f06c7c3aad4fac95f0 | | 66308 | |
| 0xfbb1b73c4f0bda4f67dca266ce6ef42f520fbb98 | Bittrex 1 | 57565 | DeFi |
| 0xea674fdde714fd979de3edf0f56aa9716b898ec8 | Ethermine | 15307 | Mining Pool |
| 0xed751387afae910bd0d2fbf75e7cd7cf60eb6abf | | 8209 | |
| 0x61c808d82a3ac53231750dadc13c777b59310bd9 | F2Pool | 8065 | Mining Pool |
| 0x5a0b54d5dc17e0aadc383d2db43b0a0d3e029c4c | Spark Pool | 6245 | Mining Pool |
| 0xe4f7a546b4ab8b0719ac14ca80871ba2dd252e87 | | 4013 | |
| 0xddd120c195b7d4975a516a0cd01df6af90e7bab7 | | 3395 | |
| 0x9c90bc6d0dd0f1ddcde0edf3b79037b50b36840b | | 3211 | |
| 0x9e65dcfdece46da8e70ae551219e8be7a676d0f4 | | 2272 | |
| 0xd91244bd83c88741b7ff8563e4482078491a1e61 | | 2231 | |
| 0x08cbce4938c2e4dc9f18176efad49abceab276e1 | | 1666 | |
| 0x36f4bfc9f49dc5d4b2d10c4a48a6b30128bd79bc | | 1586 | |
| 0xeb92130abc574b8305af10c1eaa0622862aac1af | | 1556 | |
| 0x1b126cac9caa133a0c3bb0873477b574f6f55e8e | | 1283 | |

Note: This table reports 20 senders with highest frequency in the filtered transactions. If the sender has a public name, usually, the address belongs to some mining pool or DeFi. The entity is given. Public names are accessed on *Etherscan.io*.

**Table 5. Descriptive statistics of bribing proxies**

| | Overall | | | Before | | | After | | |
|---|---|---|---|---|---|---|---|---|---|
| | Benchmark | A | B | Benchmark | A | B | Benchmark | A | B |
| **Mean** | 76.65 | 4858.84 | 4803.37 | 85.45 | 5406.77 | 5342.75 | 36.77 | 2376.11 | 2359.40 |
| **Median** | 4.57 | 17.66 | 10.86 | 5.31 | 20.85 | 12.74 | 1.63 | 8.68 | 5.16 |
| **Max** | 19708.70 | 1361047.61 | 1361045.94 | 19708.70 | 1361047.61 | 1361045.94 | 1860.43 | 468107.14 | 468107.14 |
| **Min** | 0.04 | 0.10 | 0.00 | 0.09 | 0.22 | 0.00 | 0.04 | 0.10 | 0.00 |
| **Std** | 631.33 | 53103.51 | 53103.10 | 693.59 | 56664.22 | 56664.10 | 152.82 | 32378.40 | 32379.26 |

**Table 6. Ether (ETH) and Bitcoin (BTC)**

| Panel A: ETH price | (1) | (2) | (3) | (4) | (5) | (6) |
|---|---|---|---|---|---|---|
| Benchmark | -0.20 (-0.76) | | | -0.42** (-2.41) | | |
| A | | -0.41* (-1.80) | | | -0.34** (-2.26) | |
| B | | | -0.41* (-1.79) | | | -0.33** (-2.24) |
| Post | 1.43*** (6.00) | 0.66** (2.19) | 0.66** (2.18) | 0.98*** (6.2) | 0.52*** (2.62) | 0.52*** (2.60) |
| Active | | | | 1.10*** (29.31) | 1.12*** (29.31) | 1.12*** (29.42) |
| BlockCnt | | | | -0.89*** (-3.22) | -0.90*** (-3.22) | -0.90*** (-3.22) |
| BlockTime | | | | -0.72** (-2.42) | -0.73** (-2.41) | -0.73** (-2.41) |
| AvgFeeUsd | | | | 0.04 (0.49) | 0.02 (0.20) | 0.02 (0.20) |
| N | 1156 | 1156 | 1156 | 1156 | 1156 | 1156 |
| Adj. R-sq | 0.03 | 0.01 | 0.00 | 0.58 | 0.57 | 0.57 |

| Panel B: daily return of ETH | (1) | (2) | (3) | (4) | (5) | (6) |
|---|---|---|---|---|---|---|
| Benchmark | 0.00 (0.06) | | | 0.00 (0.00) | | |
| A | | 0.19*** (3.71) | | | 0.19*** (3.72) | |
| B | | | 0.19*** (3.71) | | | 0.19*** (3.73) |
| Post | 0.06 (1.04) | -0.05 (-0.71) | -0.05 (-0.71) | 0.05 (0.99) | -0.05 (-0.76) | -0.05 (-0.76) |
| Active | | | | 0.00 (-0.06) | 0.00 (-0.02) | 0.00 (-0.02) |
| BlockCnt | | | | 0.15 (1.61) | 0.14 (1.54) | 0.14 (1.54) |
| BlockTime | | | | 0.16 (1.55) | 0.15 (1.48) | 0.15 (1.48) |
| AvgFeeUsd | | | | 0.02 (0.63) | 0.02 (0.71) | 0.02 (0.71) |
| N | 1156 | 1156 | 1156 | 1156 | 1156 | 1156 |
| Adj. R-sq | 0.00 | 0.01 | 0.01 | 0.01 | 0.01 | 0.01 |

| Panel C: ETH volume | (1) | (2) | (3) | (4) | (5) | (6) |
|---|---|---|---|---|---|---|
| Benchmark | 0.04 (0.41) | | | -0.03 (-0.49) | | |
| A | | -0.02 (-0.23) | | | 0.04 (0.68) | |
| B | | | -0.02 (-0.23) | | | 0.04 (0.49) |
| Post | 0.13 (1.46) | 0.13 (1.21) | 0.13 (1.20) | 0.02 (0.39) | 0.04 (0.58) | 0.04 (0.58) |
| Active | | | | 0.15*** (11.20) | 0.15*** (11.25) | 0.15*** (11.25) |
| BlockCnt | | | | 0.10 (1.02) | 0.10 (0.99) | 0.10 (0.99) |
| BlockTime | | | | 0.07 (0.66) | 0.07 (0.64) | 0.07 (0.64) |
| AvgFeeUsd | | | | 0.54*** (19.84) | 0.54*** (19.83) | 0.54*** (19.83) |
| N | 1156 | 1156 | 1156 | 1156 | 1156 | 1156 |
| Adj. R-sq | 0.00 | 0.00 | 0.00 | 0.56 | 0.56 | 0.56 |

| Panel D: market cap of ETH | (1) | (2) | (3) | (4) | (5) | (6) |
|---|---|---|---|---|---|---|
| Benchmark | -0.21 (-0.78) | | | -0.42** (-2.41) | | |
| A | | -0.41* (-1.80) | | | -0.34** (-2.27) | |
| B | | | -0.41* (-1.80) | | | -0.34** (-2.24) |
| Post | 1.42*** (6.00) | 0.66** (2.21) | 0.66** (2.19) | 0.98*** (6.21) | 0.52*** (2.64) | 0.52*** (2.63) |
| Active | | | | 1.10*** (29.20) | 1.12*** (29.32) | 1.12*** (29.31) |
| BlockCnt | | | | -0.88*** (-3.20) | -0.90*** (-3.21) | -0.90*** (-3.21) |
| BlockTime | | | | -0.71** (-2.40) | -0.72** (-2.40) | -0.72** (-2.40) |
| AvgFeeUsd | | | | 0.01 (0.09) | -0.01 (-0.19) | -0.01 (-0.19) |
| N | 1156 | 1156 | 1156 | 1156 | 1156 | 1156 |
| Adj. R-sq | 0.03 | 0.00 | 0.00 | 0.57 | 0.56 | 0.56 |

| Panel E: BTC price | (1) | (2) | (3) | (4) | (5) | (6) |
|---|---|---|---|---|---|---|
| Benchmark | -0.11 (-0.42) | | | -0.30 (-1.42) | | |
| A | | -0.42* (-1.83) | | | -0.46*** (-2.51) | |
| B | | | -0.42* (-1.83) | | | -0.46** (-2.50) |
| Post | 1.24*** (5.07) | 0.53* (1.72) | 0.53* (1.71) | 1.26*** (6.57) | 0.53** (2.19) | 0.53** (2.17) |
| Active | | | | 0.64*** (13.60) | 0.65*** (13.59) | 0.65*** (13.59) |
| BlockCnt | | | | 0.18 (0.82) | 0.19 (0.83) | 0.18 (0.83) |
| BlockTime | | | | 0.36 (1.11) | 0.34 (1.02) | 0.34 (1.02) |
| AvgFeeUsd | | | | 0.79*** (10.40) | 0.78*** (10.03) | 0.78*** (10.03) |
| N | 1156 | 1156 | 1156 | 1156 | 1156 | 1156 |
| Adj. R-sq | 0.02 | 0.00 | 0.00 | 0.41 | 0.39 | 0.39 |

| Panel F: daily return of BTC | (1) | (2) | (3) | (4) | (5) | (6) |
|---|---|---|---|---|---|---|
| Benchmark | -0.01 (-0.10) | | | 0.00 (-0.07) | | |
| A | | 0.07 (1.37) | | | 0.07 (1.32) | |
| B | | | 0.07 (1.37) | | | 0.07 (1.32) |
| Post | 0.04 (0.84) | -0.06 (-0.90) | -0.06 (-0.91) | 0.03 (0.65) | -0.06 (-0.89) | -0.06 (-0.90) |
| Active | | | | 0.03*** (2.51) | 0.03*** (2.51) | 0.03*** (2.51) |
| BlockCnt | | | | 0.09 (1.49) | 0.09 (1.51) | 0.09 (1.51) |
| BlockTime | | | | 0.10 (1.12) | 0.10 (1.13) | 0.10 (1.13) |
| AvgFeeUsd | | | | -0.02 (-1.13) | -0.02 (-1.13) | -0.02 (-1.13) |
| N | 1156 | 1156 | 1156 | 1156 | 1156 | 1156 |
| Adj. R-sq | 0.00 | 0.00 | 0.00 | 0.01 | 0.01 | 0.01 |

| Panel G: BTC volume | (1) | (2) | (3) | (4) | (5) | (6) |
|---|---|---|---|---|---|---|
| Benchmark | 0.01 (0.08) | | | -0.08 (-1.17) | | |
| A | | 0.02 (0.26) | | | 0.02 (0.32) | |
| B | | | 0.02 (0.26) | | | 0.02 (0.34) |
| Post | 0.07 (0.79) | 0.09 (0.80) | 0.09 (0.80) | 0.10 (1.50) | 0.09 (1.14) | 0.09 (1.14) |
| Active | | | | 0.16*** (10.10) | 0.16*** (10.09) | 0.16*** (10.09) |
| BlockCnt | | | | 0.07 (1.00) | 0.07 (0.99) | 0.07 (0.99) |
| BlockTime | | | | 0.18 (1.62) | 0.17 (1.57) | 0.17 (1.57) |
| AvgFeeUsd | | | | 0.41*** (16.06) | 0.41*** (16.03) | 0.41*** (16.03) |
| N | 1156 | 1156 | 1156 | 1156 | 1156 | 1156 |
| Adj. R-sq | 0.00 | 0.00 | 0.00 | 0.44 | 0.44 | 0.44 |

| Panel H: market cap of BTC | (1) | (2) | (3) | (4) | (5) | (6) |
|---|---|---|---|---|---|---|
| Benchmark | -0.11 (0.06) | | | -0.30 (-1.41) | | |
| A | | -0.42* (-1.83) | | | -0.46*** (-2.52) | |
| B | | | -0.42* (-1.83) | | | -0.46** (-2.51) |
| Post | 1.24*** (5.10) | 0.53* (1.72) | 0.53* (1.72) | 1.25*** (6.57) | 0.53** (2.18) | 0.52** (2.17) |
| Active | | | | 0.64*** (13.62) | 0.65*** (13.62) | 0.65*** (13.62) |
| BlockCnt | | | | 0.18 (0.80) | 0.18 (0.81) | 0.18 (0.81) |
| BlockTime | | | | 0.37 (1.12) | 0.34 (1.03) | 0.34 (1.03) |
| AvgFeeUsd | | | | 0.77*** (10.13) | 0.75*** (9.75) | 0.75*** (9.75) |
| N | 1156 | 1156 | 1156 | 1156 | 1156 | 1156 |
| Adj. R-sq | 0.02 | 0.00 | 0.00 | 0.40 | 0.38 | 0.38 |

Note: This table reports regression results. In Columns (1) - (3) of each panel, we run the regression model: $token_{i,t} = \beta_0 + \beta_1 bribing_t + \beta_2 post_t \times bribing_t + \varepsilon_{i,t}$, using proxy benchmark, A and B, respectively. In Columns (4) – (6), we consider control variables in the regression model: $token_{i,t} = \beta_0 + \beta_1 bribing_t + \beta_2 control_{i,t} + \beta_3 post_t \times bribing_t + \varepsilon_{i,t}$, and the independent variable $bribing$ is proxy benchmark, A and B, respectively. T-statistics are reported in parentheses. *, **, and *** denote significance levels at the 10%, 5%, and 1% levels based on the standard t-statistics.

**Table 7. Transaction statistics of Ethereum**

| Panel A: TxnVol | (1) | (2) | (3) | (4) | (5) | (6) |
|---|---|---|---|---|---|---|
| Benchmark | 0.08 | | | 0.04 | | |
| | (1.20) | | | (0.74) | | |
| **A** | | 0.06 | | | **0.08*** | |
| | | (0.98) | | | **(1.83)** | |
| **B** | | | 0.06 | | | **0.08*** |
| | | | (0.96) | | | **(1.82)** |
| Post | -0.04 | -0.03 | -0.03 | -0.10 | -0.08 | -0.08 |
| | (-0.62) | (-0.36) | (-0.36) | (-2.00) | (-1.24) | (-1.23) |
| **Active** | | | | **0.09*** | **0.09*** | **0.09*** |
| | | | | **(7.97)** | **(7.84)** | **(7.84)** |
| **BlockCnt** | | | | **0.20**** | **0.19**** | **0.19**** |
| | | | | **(2.25)** | **(2.25)** | **(2.25)** |
| **BlockTime** | | | | **0.19**** | **0.19**** | **0.19**** |
| | | | | **(2.06)** | **(2.05)** | **(2.05)** |
| **AvgFeeUsd** | | | | **0.28*** | **0.28*** | **0.28*** |
| | | | | **(11.89)** | **(12.01)** | **(12.01)** |
| N | 1156 | 1156 | 1156 | 1156 | 1156 | 1156 |
| Adj. R-sq | 0.00 | 0.00 | 0.00 | 0.35 | 0.35 | 0.35 |

| Panel B: TxnVolUsd | (1) | (2) | (3) | (4) | (5) | (6) |
|---|---|---|---|---|---|---|
| Benchmark | 0.02 | | | -0.02 | | |
| | (0.39) | | | (-0.60) | | |
| A | | -0.05 | | | -0.03 | |
| | | (-1.19) | | | (-0.94) | |
| B | | | -0.05 | | | -0.03 |
| | | | (-1.19) | | | (-0.94) |
| **Post** | **0.12*** | 0.07 | 0.07 | **0.05*** | 0.03 | 0.03 |
| | **(2.62)** | (1.19) | (1.19) | **(1.72)** | (0.73) | (0.72) |
| **Active** | | | | **0.14*** | **0.14*** | **0.14*** |
| | | | | **(18.52)** | **(18.68)** | **(18.68)** |
| **BlockCnt** | | | | **-0.19*** | **-0.19*** | **-0.19*** |
| | | | | **(-3.46)** | **(-3.46)** | **(-3.46)** |
| **BlockTime** | | | | **-0.17*** | **-0.17*** | **-0.17*** |
| | | | | **(-2.90)** | **(-2.90)** | **(-2.90)** |
| **AvgFeeUsd** | | | | **0.21*** | **0.21*** | **0.21*** |
| | | | | **(14.15)** | **(14.05)** | **(14.05)** |
| N | 1156 | 1156 | 1156 | 1156 | 1156 | 1156 |
| Adj. R-sq | 0.01 | 0.00 | 0.00 | 0.56 | 0.56 | 0.56 |

| Panel C: TxnCnt | (1) | (2) | (3) | (4) | (5) | (6) |
|---|---|---|---|---|---|---|
| Benchmark | 0.13 | | | -0.06 | | |
| | (0.64) | | | (-0.97) | | |
| A | | -0.09 | | | 0.01 | |
| | | (-0.48) | | | (0.23) | |
| B | | | -0.08 | | | 0.01 |
| | | | (-0.49) | | | (0.24) |
| **Post** | **0.55*** | 0.21 | 0.21 | **0.16*** | 0.05 | 0.05 |
| | **(3.09)** | (0.95) | (0.95) | **(2.81)** | (0.75) | (0.74) |
| **Active** | | | | **0.88*** | **0.88*** | **0.88*** |
| | | | | **(63.86)** | **(64.03)** | **(64.03)** |
| BlockCnt | | | | 0.11 | 0.11 | 0.11 |
| | | | | (1.09) | (1.05) | (1.05) |
| BlockTime | | | | -0.01 | -0.01 | -0.01 |
| | | | | (-0.05) | (-0.09) | (-0.09) |
| **AvgFeeUsd** | | | | **0.21*** | **0.20*** | **0.20*** |
| | | | | **(7.49)** | **(7.37)** | **(7.37)** |
| N | 1156 | 1156 | 1156 | 1156 | 1156 | 1156 |
| Adj. R-sq | 0.01 | 0.00 | 0.00 | 0.90 | 0.90 | 0.90 |

Note: This table reports regression results. In Columns (1) - (3) of each panel, we run the regression model: $chain_{i,t} = \beta_0 + \beta_1 bribing_t + \beta_2 post_t \times bribing_t + \varepsilon_{i,t}$, using proxy benchmark, A and B, respectively. In Columns (4) – (6), we consider control variables in the regression model: $chain_{i,t} = \beta_0 + \beta_1 bribing_t + \beta_2 control_{i,t} + \beta_3 post_t \times bribing_t + \varepsilon_{i,t}$, and the independent variable $bribing$ is proxy benchmark, A and B, respectively. T-statistics are reported in parentheses. *, **, and *** denote significance levels at the 10%, 5%, and 1% levels based on the standard t-statistics.

**Table 8. Transaction statistics of Bitcoin**

| **Panel A: TxnVol** | | | | | | |
|---|---|---|---|---|---|---|
| | (1) | (2) | (3) | (4) | (5) | (6) |
| Benchmark | -0.07 | | | -0.07 | | |
| | (-1.12) | | | (-1.15) | | |
| A | | -0.06 | | | -0.07 | |
| | | (-1.10) | | | (-1.51) | |
| B | | | -0.06 | | | -0.07 |
| | | | (-1.09) | | | (-1.51) |
| **Post** | **0.15**** | 0.10 | 0.10 | **0.13**** | 0.08 | 0.08 |
| | **(2.72)** | (1.46) | (1.46) | **(2.54)** | (1.33) | (1.32) |
| **Active** | | | | **0.15**** | **0.15**** | **0.15**** |
| | | | | **(12.18)** | **(12.24)** | **(12.24)** |
| BlockCnt | | | | -0.08 | -0.08 | -0.08 |
| | | | | (-1.28) | (-1.31) | (-1.31) |
| BlockTime | | | | -0.02 | -0.03 | -0.03 |
| | | | | (-0.25) | (-0.33) | (-0.33) |
| **AvgFeeUsd** | | | | **-0.13**** | **-0.13**** | **-0.13**** |
| | | | | **(-6.51)** | **(-6.60)** | **(-6.60)** |
| N | 1156 | 1156 | 1156 | 1156 | 1156 | 1156 |
| Adj. R-sq | 0.01 | 0.00 | 0.00 | 0.12 | 0.12 | 0.12 |
| **Panel B: TxnVolUsd** | | | | | | |
| | (1) | (2) | (3) | (4) | (5) | (6) |
| Benchmark | -0.11 | | | -0.13 | | |
| | (-1.15) | | | (-1.38) | | |
| A | | -0.13 | | | **-0.16**** | |
| | | (-1.56) | | | **(-2.00)** | |
| B | | | -0.13 | | | **-0.16**** |
| | | | (-1.55) | | | **(-1.99)** |
| **Post** | **0.44**** | **0.21*** | **0.21*** | **0.41**** | **0.20*** | **0.20*** |
| | **(4.87)** | **(1.89)** | **(1.89)** | **(4.99)** | **(1.90)** | **(1.89)** |
| **Active** | | | | **0.25**** | **0.26**** | **0.26**** |
| | | | | **(12.51)** | **(12.55)** | **(12.56)** |
| BlockCnt | | | | -0.08 | -0.08 | -0.08 |
| | | | | (-0.80) | (-0.79) | (-0.79) |
| BlockTime | | | | -0.03 | -0.04 | -0.04 |
| | | | | (-0.20) | (-0.27) | (-0.27) |
| **AvgFeeUsd** | | | | **-0.07**** | **-0.07**** | **-0.07**** |
| | | | | **(-2.05)** | **(-2.20)** | **(-2.21)** |
| N | 1156 | 1156 | 1156 | 1156 | 1156 | 1156 |
| Adj. R-sq | 0.02 | 0.00 | 0.00 | 0.17 | 0.15 | 0.15 |
| **Panel C: TxnCnt** | | | | | | |
| | (1) | (2) | (3) | (4) | (5) | (6) |
| Benchmark | -0.04 | | | 0.10 | | |
| | (-0.33) | | | (0.81) | | |
| A | | **0.28**** | | | **0.25**** | |
| | | **(2.55)** | | | **(2.52)** | |
| B | | | **0.28**** | | | **0.25**** |
| | | | **(2.56)** | | | **(2.52)** |
| **Post** | **-0.33**** | -0.18 | -0.17 | **-0.44**** | -0.17 | -0.17 |
| | **(-2.85)** | (-1.21) | (-1.20) | **(-4.14)** | (-1.29) | (-1.27) |
| **Active** | | | | **0.15**** | **0.14**** | **0.14**** |
| | | | | **(5.58)** | **(5.39)** | **(5.39)** |
| BlockCnt | | | | -0.14 | -0.14 | -0.14 |
| | | | | (-1.15) | (-1.14) | (-1.14) |
| **BlockTime** | | | | **-0.78**** | **-0.77**** | **-0.77**** |
| | | | | **(-4.33)** | **(-4.26)** | **(-4.26)** |
| **AvgFeeUsd** | | | | **-0.17**** | **-0.16**** | **-0.16**** |
| | | | | **(-3.98)** | **(-3.80)** | **(-3.80)** |
| N | 1156 | 1156 | 1156 | 1156 | 1156 | 1156 |
| Adj. R-sq | 0.01 | 0.00 | 0.00 | 0.18 | 0.17 | 0.17 |

Note: This table reports regression results. In Columns (1) - (3) of each panel, we run the regression model: $chain_{i,t} = \beta_0 + \beta_1 bribing_t + \beta_2 post_t \times bribing_t + \varepsilon_{i,t}$, using proxy benchmark, A and B, respectively. In Columns (4) – (6), we consider control variables in the regression model: $chain_{i,t} = \beta_0 + \beta_1 bribing_t + \beta_2 control_{i,t} + \beta_3 post_t \times bribing_t + \varepsilon_{i,t}$, and the independent variable $bribing$ is proxy benchmark, A and B, respectively. T-statistics are reported in parentheses. *, **, and *** denote significance levels at the 10%, 5%, and 1% levels based on the standard t-statistics.

**Table 9. Network factors of Ethereum**

| Panel A: Unique | | | | | | |
|---|---|---|---|---|---|---|
| | (1) | (2) | (3) | (4) | (5) | (6) |
| Benchmark | -0.09 | | | 0.01 | | |
| | (0.36) | | | (0.09) | | |
| A | | -0.43** | | | -0.14 | |
| | | (-1.97) | | | (-1.47) | |
| B | | | -0.43** | | | -0.15 |
| | | | (-1.97) | | | (-1.48) |
| Post | **0.90*** | 0.46 | 0.46 | **-0.21**** | -0.05 | -0.04 |
| | **(3.87)** | (1.58) | (1.58) | **(-2.01)** | (-0.35) | (-0.34) |
| Price | | | | **0.78*** | **0.78*** | **0.78*** |
| | | | | **(52.62)** | **(53.01)** | **(53.01)** |
| TxnVol | | | | **0.58*** | **0.59*** | **0.59*** |
| | | | | **(9.59)** | **(9.68)** | **(9.68)** |
| BlockCnt | | | | **1.58*** | **1.58*** | **1.58*** |
| | | | | **(8.79)** | **(8.82)** | **(8.82)** |
| BlockTime | | | | **1.40*** | **1.41*** | **1.41*** |
| | | | | **(7.17)** | **(7.20)** | **(7.20)** |
| AvgFeeUsd | | | | **-0.15*** | **-0.15*** | **-0.15*** |
| | | | | **(-2.88)** | **(-2.83)** | **(-2.83)** |
| N | 1156 | 1156 | 1156 | 1156 | 1156 | 1156 |
| Adj. R-sq | 0.01 | 0.00 | 0.00 | 0.81 | 0.81 | 0.81 |
| **Panel B: New** | | | | | | |
| | (1) | (2) | (3) | (4) | (5) | (6) |
| Benchmark | 0.17 | | | 0.10 | | |
| | (1.26) | | | (0.98) | | |
| A | | 0.03 | | | 0.07 | |
| | | (0.24) | | | (0.78) | |
| B | | | 0.03 | | | 0.07 |
| | | | (0.23) | | | (0.77) |
| Post | -0.07 | -0.05 | -0.05 | -0.15 | -0.14 | -0.14 |
| | (-0.56) | (-0.36) | (-0.35) | (-1.61) | (-1.19) | (-1.18) |
| Price | | | | **0.04*** | **0.03*** | **0.03*** |
| | | | | **(2.74)** | **(2.55)** | **(2.54)** |
| TxnVol | | | | **0.36*** | **0.36*** | **0.36*** |
| | | | | **(6.54)** | **(6.52)** | **(6.52)** |
| BlockCnt | | | | **0.77*** | **0.77*** | **0.77*** |
| | | | | **(4.73)** | **(4.74)** | **(4.74)** |
| BlockTime | | | | **0.64*** | **0.64*** | **0.64*** |
| | | | | **(3.59)** | **(3.60)** | **(3.60)** |
| AvgFeeUsd | | | | **0.48*** | **0.49*** | **0.49*** |
| | | | | **(10.26)** | **(10.38)** | **(10.38)** |
| N | 1156 | 1156 | 1156 | 1156 | 1156 | 1156 |
| Adj. R-sq | 0.00 | 0.00 | 0.00 | 0.37 | 0.37 | 0.37 |
| **Panel C: Active** | | | | | | |
| | (1) | (2) | (3) | (4) | (5) | (6) |
| **Benchmark** | 0.20 | | | **0.19**** | | |
| | (1.05) | | | **(2.00)** | | |
| A | | -0.08 | | | 0.09 | |
| | | (-0.47) | | | (1.15) | |
| B | | | -0.08 | | | 0.09 |
| | | | (-0.48) | | | (1.13) |
| Post | **0.42**** | 0.14 | 0.14 | **-0.18**** | -0.17 | -0.17 |
| | **(2.43)** | (0.66) | (0.66) | **(-2.04)** | (-1.59) | (-1.58) |
| Price | | | | **0.40*** | **0.40*** | **0.40*** |
| | | | | **(33.07)** | **(33.21)** | **(33.21)** |
| TxnVol | | | | **0.72*** | **0.72*** | **0.72*** |
| | | | | **(14.35)** | **(14.31)** | **(14.31)** |
| BlockCnt | | | | **1.07*** | **1.07*** | **1.07*** |
| | | | | **(7.23)** | **(7.24)** | **(7.24)** |
| BlockTime | | | | **0.88*** | **0.88*** | **0.88*** |
| | | | | **(5.46)** | **(5.47)** | **(5.47)** |
| AvgFeeUsd | | | | **0.35*** | **0.36*** | **0.36*** |
| | | | | **(8.34)** | **(8.50)** | **(8.50)** |
| N | 1156 | 1156 | 1156 | 1156 | 1156 | 1156 |
| Adj. R-sq | 0.01 | 0.00 | 0.00 | 0.76 | 0.76 | 0.76 |
| **Panel D: Active.Ratio** | | | | | | |
| | (1) | (2) | (3) | (4) | (5) | (6) |
| **Benchmark** | **0.25**** | | | 0.18 | | |
| | **(2.13)** | | | (1.63) | | |
| A | | **0.40*** | | | **0.35*** | |
| | | **(3.98)** | | | **(3.87)** | |
| B | | | **0.40*** | | | **0.35*** |
| | | | **(3.97)** | | | **(3.85)** |
| Post | **-0.26**** | **-0.25*** | **-0.25*** | -0.04 | -0.19 | -0.19 |
| | **(-2.41)** | **(-1.90)** | **(-1.89)** | (-0.42) | (-1.54) | (-1.54) |
| Price | | | | **-0.16*** | **-0.16*** | **-0.16*** |
| | | | | **(-11.63)** | **(-11.81)** | **(-11.81)** |
| TxnVol | | | | **0.28*** | **0.27*** | **0.27*** |
| | | | | **(4.94)** | **(4.80)** | **(4.80)** |
| BlockCnt | | | | -0.16 | -0.17 | -0.17 |
| | | | | (-0.97) | (-1.01) | (-1.01) |
| BlockTime | | | | -0.27 | -0.28 | -0.28 |

|  | (1) | (2) | (3) | (4) | (5) | (6) |
|---|---|---|---|---|---|---|
|  |  |  |  | (-1.47) | (-1.52) | (-1.52) |
| AvgFeeUsd |  |  |  | 0.35*** | 0.35*** | 0.35*** |
|  |  |  |  | (7.12) | (7.34) | (7.34) |
| N | 1156 | 1156 | 1156 | 1156 | 1156 | 1156 |
| Adj. R-sq | 0.01 | 0.01 | 0.01 | 0.17 | 0.18 | 0.18 |

Note: This table reports regression results. In Columns (1) - (3) of each panel, we run the univariate regression model: $network_{i,t} = \beta_0 + \beta_1 bribing_t + \beta_2 post_t \times bribing_t + \varepsilon_{i,t}$, using proxy benchmark, A and B, respectively. In Columns (4) – (6), we consider control variables in the regression model: $network_{i,t} = \beta_0 + \beta_1 bribing_t + \beta_2 control_{i,t} + \beta_3 post_t \times bribing_t + \varepsilon_{i,t}$, and the independent variable $bribing$ is proxy benchmark, A and B, respectively. T-statistics are reported in parentheses. *, **, and *** denote significance levels at the 10%, 5%, and 1% levels based on the standard t-statistics.

**Table 10. Network factors of Bitcoin**

| **Panel A: Unique** | | | | | | |
|---|---|---|---|---|---|---|
|  | (1) | (2) | (3) | (4) | (5) | (6) |
| Benchmark | -0.07 |  |  | 0.00 |  |  |
|  | (-0.27) |  |  | (0.00) |  |  |
| A |  | -0.49** |  |  | -0.08 |  |
|  |  | (-2.08) |  |  | (-0.75) |  |
| B |  |  | -0.49** |  |  | -0.08 |
|  |  |  | (-2.08) |  |  | (-0.76) |
| Post | 1.04*** | 0.56* | 0.56* | -0.12 | 0.04 | 0.04 |
|  | (4.21) | (1.82) | (1.81) | (-1.06) | (0.26) | (0.26) |
| Price |  |  |  | 0.93*** | 0.93*** | 0.93*** |
|  |  |  |  | (49.81) | (50.23) | (50.23) |
| TxnVol |  |  |  | 0.20*** | 0.20*** | 0.20*** |
|  |  |  |  | (2.79) | (2.81) | (2.81) |
| BlockCnt |  |  |  | 0.21* | 0.21* | 0.21* |
|  |  |  |  | (1.67) | (1.65) | (1.65) |
| BlockTime |  |  |  | 0.72*** | 0.72*** | 0.72*** |
|  |  |  |  | (3.85) | (3.84) | (3.84) |
| AvgFeeUsd |  |  |  | -0.30*** | -0.29*** | -0.29*** |
|  |  |  |  | (-6.82) | (-6.71) | (-6.71) |
| N | 1156 | 1156 | 1156 | 1156 | 1156 | 1156 |
| Adj. R-sq | 0.01 | 0.00 | 0.00 | 0.81 | 0.80 | 0.80 |
| **Panel B: New** | | | | | | |
|  | (1) | (2) | (3) | (4) | (5) | (6) |
| Benchmark | 0.04 |  |  | 0.08 |  |  |
|  | (0.28) |  |  | (0.66) |  |  |
| A |  | 0.16 |  |  | 0.18* |  |
|  |  | (1.32) |  |  | (1.87) |  |
| B |  |  | 0.16 |  |  | 0.18* |
|  |  |  | (1.32) |  |  | (1.86) |
| Post | -0.05 | -0.01 | -0.01 | -0.11 | -0.02 | -0.02 |
|  | (-0.37) | (-0.08) | (-0.08) | (-1.05) | (-0.16) | (-0.15) |
| Price |  |  |  | 0.00 | 0.00 | 0.00 |
|  |  |  |  | (0.16) | (0.05) | (0.05) |
| TxnVol |  |  |  | 0.38*** | 0.38*** | 0.38*** |
|  |  |  |  | (5.70) | (5.73) | (5.73) |
| BlockCnt |  |  |  | -0.31*** | -0.30*** | -0.30*** |
|  |  |  |  | (-2.60) | (-2.54) | (-2.54) |
| BlockTime |  |  |  | -0.97*** | -0.95*** | -0.95*** |
|  |  |  |  | (-5.48) | (-5.42) | (-5.41) |
| AvgFeeUsd |  |  |  | 0.68*** | 0.68*** | 0.68*** |
|  |  |  |  | (16.62) | (16.85) | (16.85) |
| N | 1156 | 1156 | 1156 | 1156 | 1156 | 1156 |
| Adj. R-sq | 0.00 | 0.00 | 0.00 | 0.32 | 0.32 | 0.32 |
| **Panel C: Active** | | | | | | |
|  | (1) | (2) | (3) | (4) | (5) | (6) |
| Benchmark | 0.05 |  |  | 0.10 |  |  |
|  | (0.30) |  |  | (0.80) |  |  |
| A |  | 0.09 |  |  | 0.18* |  |
|  |  | (0.61) |  |  | (1.79) |  |
| B |  |  | 0.09 |  |  | 0.18* |
|  |  |  | (0.60) |  |  | (1.78) |
| Post | 0.11 | 0.04 | 0.04 | -0.16 | -0.04 | -0.04 |
|  | (0.72) | (0.20) | (0.20) | (-1.41) | (-0.33) | (-0.32) |
| Price |  |  |  | 0.16*** | 0.15*** | 0.15*** |
|  |  |  |  | (8.40) | (8.34) | (8.34) |
| TxnVol |  |  |  | 0.43*** | 0.43*** | 0.43*** |
|  |  |  |  | (6.11) | (6.14) | (6.14) |
| BlockCnt |  |  |  | -0.09 | -0.08 | -0.08 |
|  |  |  |  | (-0.73) | (-0.66) | (-0.66) |
| BlockTime |  |  |  | -0.74*** | -0.72*** | -0.72*** |
|  |  |  |  | (-3.97) | (-3.89) | (-3.89) |
| AvgFeeUsd |  |  |  | 0.76*** | 0.77*** | 0.77*** |
|  |  |  |  | (17.68) | (17.94) | (17.94) |
| N | 1156 | 1156 | 1156 | 1156 | 1156 | 1156 |

|  | | (1) | (2) | (3) | (4) | (5) | (6) |
|---|---|---|---|---|---|---|---|
| Adj. R-sq | | 0.00 | 0.00 | 0.00 | 0.49 | 0.49 | 0.49 |
| **Panel D: Active.Ratio** | | | | | | | |
|  | | (1) | (2) | (3) | (4) | (5) | (6) |
| Benchmark | | 0.09 | | | 0.10 | | |
|  | | (0.64) | | | (0.88) | | |
| A | | | 0.40*** | | | 0.28*** | |
|  | | | (3.21) | | | (2.84) | |
| B | | | | 0.40*** | | | 0.28*** |
|  | | | | (3.21) | | | (2.83) |
| Post | | -0.43*** | -0.28* | -0.28* | -0.08 | -0.08 | -0.08 |
|  | | (-3.20) | (-1.67) | (-1.66) | (-0.70) | (-0.62) | (-0.62) |
| Price | | | | | -0.35*** | -0.35*** | -0.35*** |
|  | | | | | (-19.49) | (-19.79) | (-19.79) |
| TxnVol | | | | | 0.39*** | 0.39*** | 0.39*** |
|  | | | | | (5.71) | (5.74) | (5.74) |
| BlockCnt | | | | | -0.10 | -0.09 | -0.09 |
|  | | | | | (-0.83) | (-0.74) | (-0.74) |
| BlockTime | | | | | -0.94*** | -0.92*** | -0.92*** |
|  | | | | | (-5.19) | (-5.10) | (-5.10) |
| AvgFeeUsd | | | | | 0.84*** | 0.85*** | 0.85*** |
|  | | | | | (20.16) | (20.39) | (20.39) |
| N | | 1156 | 1156 | 1156 | 1156 | 1156 | 1156 |
| Adj. R-sq | | 0.01 | 0.01 | 0.01 | 0.37 | 0.38 | 0.38 |

Note: This table reports regression results. In Columns (1) - (3) of each panel, we run the univariate regression model: $network_{i,t} = \beta_0 + \beta_1 bribing_t + \beta_2 post_t \times bribing_t + \varepsilon_{i,t}$, using proxy benchmark, A and B, respectively. In Columns (4) – (6), we consider control variables in the regression model: $network_{i,t} = \beta_0 + \beta_1 bribing_t + \beta_2 control_{i,t} + \beta_3 post_t \times bribing_t + \varepsilon_{i,t}$, and the independent variable $bribing$ is proxy benchmark, A and B, respectively. T-statistics are reported in parentheses. *, **, and *** denote significance levels at the 10%, 5%, and 1% levels based on the standard t-statistics.

### Table 11. Stock markets

**Panel A: S&P500**

|  | (1) | (2) | (3) | (4) | (5) | (6) | (7) | (8) | (9) |
|---|---|---|---|---|---|---|---|---|---|
| Benchmark | -0.03 | | | 0.07 | | | 0.04 | | |
|  | (0.38) | | | (0.59) | | | (0.49) | | |
| A | | -0.14* | | | -0.14 | | | 0.08 | |
|  | | (-1.74) | | | (-1.24) | | | (1.07) | |
| B | | | -0.14* | | | -0.14 | | | 0.08 |
|  | | | (-1.73) | | | (-1.26) | | | (1.07) |
| Post | 0.03 | 0.05 | 0.05 | -0.14 | 0.00 | 0.00 | -0.01 | 0.05 | 0.05 |
|  | (0.36) | (0.50) | (0.50) | (-1.25) | (0.01) | (0.01) | (-0.16) | (0.57) | (0.57) |
| Price | 0.80*** | 0.80*** | 0.80*** | -0.20*** | -0.20*** | -0.20*** | -0.01 | -0.01 | -0.01 |
|  | (62.50) | (63.37) | (63.38) | (-11.36) | (-11.73) | (-11.73) | (-0.70) | (-0.75) | (-0.75) |
| TxnVol | 0.36*** | 0.37*** | 0.37*** | 0.22*** | 0.23*** | 0.23*** | -0.09** | -0.09 | -0.09 |
|  | (8.36) | (8.46) | (8.46) | (3.70) | (3.80) | (3.80) | (-2.27) | (-2.31) | (-2.31) |
| BlockCnt | 0.58*** | 0.59*** | 0.59*** | 1.41*** | 1.41*** | 1.41*** | 0.01 | 0.01 | 0.01 |
|  | (3.77) | (3.79) | (3.79) | (6.55) | (6.57) | (6.57) | (0.06) | (0.05) | (0.05) |
| BlockTime | 0.51*** | 0.51*** | 0.51*** | 1.39*** | 1.40*** | 1.40*** | 0.03 | 0.03 | 0.03 |
|  | (3.00) | (3.02) | (3.02) | (5.91) | (5.93) | (5.93) | (0.20) | (0.19) | (0.19) |
| AvgFeeUsd | -0.11*** | -0.11*** | -0.11*** | -0.12** | 0.12** | 0.12** | 0.04 | 0.05 | 0.05 |
|  | (-2.70) | (-2.80) | (-2.80) | (2.10) | (2.12) | (2.12) | (1.15) | (1.19) | (1.19) |
| N | 798 | 798 | 798 | 798 | 798 | 798 | 798 | 798 | 798 |
| Adj. R-sq | 0.88 | 0.88 | 0.88 | 0.21 | 0.21 | 0.21 | 0.00 | 0.00 | 0.01 |

**Panel B: NASDAQ**

|  | (1) | (2) | (3) | (4) | (5) | (6) | (7) | (8) | (9) |
|---|---|---|---|---|---|---|---|---|---|
| Benchmark | 0.00 | | | 0.01 | | | 0.02 | | |
|  | (-0.03) | | | (0.07) | | | (0.94) | | |
| A | | -0.23** | | | -0.12 | | | 0.02 | |
|  | | (-2.14) | | | (-1.32) | | | (0.27) | |
| B | | | -0.23** | | | -0.12 | | | 0.02 |
|  | | | (-2.14) | | | (-1.32) | | | (0.26) |
| Post | 0.07 | 0.11 | 0.11 | -0.08 | -0.04 | -0.04 | 0.00 | -0.02 | -0.02 |
|  | (0.74) | (0.94) | (0.94) | (-0.90) | (-0.38) | (-0.38) | (0.07) | (-0.22) | (-0.22) |
| Price | 0.83*** | 0.83*** | 0.83*** | 0.16*** | 0.16*** | 0.16*** | -0.01 | -0.01 | -0.01 |
|  | (51.31) | (52.09) | (52.09) | (10.89) | (10.88) | (10.88) | (-0.75) | (-0.75) | (-0.75) |
| TxnVol | 0.70*** | 0.71*** | 0.71*** | 0.42*** | 0.42*** | 0.42*** | -0.01 | -0.01 | -0.01 |
|  | (12.80) | (12.94) | (12.94) | (8.54) | (8.62) | (8.62) | (-0.38) | (-0.40) | (-0.40) |
| BlockCnt | 1.43*** | 1.43*** | 1.43*** | 1.44*** | 1.44*** | 1.44*** | -0.02 | -0.02 | -0.02 |
|  | (7.25) | (7.30) | (7.30) | (8.20) | (8.22) | (8.22) | (-0.16) | (-0.14) | (-0.14) |
| BlockTime | 1.34*** | 1.35*** | 1.35*** | 1.44*** | 1.44*** | 1.44*** | -0.01 | -0.01 | -0.01 |
|  | (6.22) | (6.27) | (6.27) | (7.47) | (7.50) | (7.50) | (-0.09) | (-0.06) | (-0.06) |
| AvgFeeUsd | -0.06 | -0.07 | -0.07 | 0.18*** | 0.18*** | 0.18*** | -0.01 | -0.01 | -0.01 |
|  | (-1.16) | (-1.28) | (-1.28) | (3.85) | (3.81) | (3.81) | (-0.31) | (-0.28) | (-0.28) |
| N | 797 | 797 | 797 | 797 | 797 | 797 | 797 | 797 | 797 |
| Adj. R-sq | 0.86 | 0.86 | 0.86 | 0.48 | 0.48 | 0.48 | 0.00 | -0.01 | -0.01 |

**Panel C: N225**

|  | (1) | (2) | (3) | (4) | (5) | (6) | (7) | (8) | (9) |
|---|---|---|---|---|---|---|---|---|---|
| Benchmark | -0.05 | | | -0.02 | | | -0.23*** | | |

|  | (1) | (2) | (3) | (4) | (5) | (6) | (7) | (8) | (9) |
|---|---|---|---|---|---|---|---|---|---|
|  | (-0.49) |  |  | (-0.17) |  |  | **(-2.62)** |  |  |
| A |  | -0.19 |  |  | 0.03 |  |  | -0.07 |  |
|  |  | (-1.74) |  |  | (0.36) |  |  | (-0.82) |  |
| B |  |  | -0.19 |  |  | 0.03 |  |  | -0.07 |
|  |  |  | (-1.74) |  |  | (0.36) |  |  | (-0.79) |
| Post | -0.05 | 0.13 | 0.13 | -0.07 | 0.03 | 0.03 | 0.04 | -0.02 | -0.02 |
|  | (-0.49) | (1.06) | (1.06) | (-0.83) | (0.32) | (0.32) | (0.53) | (-0.17) | (-0.18) |
| Price | **0.60\*\*\*** | **0.60\*\*\*** | **0.60\*\*\*** | **-0.04\*\*\*** | **-0.04\*\*\*** | **-0.04\*\*\*** | 0.00 | 0.00 | 0.00 |
|  | **(34.85)** | **(35.18)** | **(35.18)** | **(-2.66)** | **(-2.84)** | **(-2.84)** | (-0.29) | (-0.16) | (-0.15) |
| TxnVol | **0.52\*\*\*** | **0.53\*\*\*** | **0.53\*\*\*** | -0.01 | -0.01 | -0.01 | -0.01 | -0.01 | -0.01 |
|  | **(9.07)** | **(9.21)** | **(9.21)** | (-0.21) | (-0.20) | (-0.20) | (-0.21) | (-0.19) | (-0.19) |
| BlockCnt | 0.33 | 0.33 | 0.33 | **0.45\*\*\*** | **0.44\*\*\*** | **0.44\*\*\*** | 0.00 | 0.00 | 0.00 |
|  | (1.58) | (1.60) | (1.60) | **(2.66)** | **(2.64)** | **(2.64)** | (0.02) | (0.02) | (0.02) |
| BlockTime | 0.33 | 0.34 | 0.34 | **0.41\*\*** | **0.41\*\*** | **0.41\*\*** | 0.02 | 0.02 | 0.02 |
|  | (1.48) | (1.51) | (1.51) | **(2.24)** | **(2.22)** | **(2.22)** | (0.09) | (0.10) | (0.10) |
| AvgFeeUsd | **0.38\*\*\*** | **0.38\*\*\*** | **0.38\*\*\*** | 0.02 | 0.03 | 0.03 | -0.02 | -0.02 | -0.02 |
|  | **(6.98)** | **(6.93)** | **(6.93)** | (0.55) | (0.59) | (0.59) | (-0.46) | (-0.55) | (-0.55) |
| N | 766 | 766 | 766 | 766 | 766 | 766 | 766 | 766 | 766 |
| Adj. R-sq | 0.77 | 0.77 | 0.77 | 0.02 | 0.02 | 0.02 | 0.00 | -0.01 | -0.01 |
| **Panel D: SSE** |  |  |  |  |  |  |  |  |  |
|  | (1) | (2) | (3) | (4) | (5) | (6) | (7) | (8) | (9) |
| Benchmark | -0.15 |  |  | -0.10 |  |  | -0.08 |  |  |
|  | (-1.33) |  |  | (-0.70) |  |  | (-0.97) |  |  |
| A |  | -0.13 |  |  | 0.01 |  |  | 0.02 |  |
|  |  | (-1.15) |  |  | (0.04) |  |  | (0.20) |  |
| B |  |  | -0.13 |  |  | 0.01 |  |  | 0.02 |
|  |  |  | (-1.13) |  |  | (0.05) |  |  | (0.22) |
| Post | -0.05 | 0.15 | 0.15 | 0.06 | **0.42\*\*\*** | **0.42\*\*\*** | -0.07 | 0.03 | 0.03 |
|  | (-0.46) | (1.18) | (1.17) | (0.47) | **(2.59)** | **(2.58)** | (-0.95) | (0.34) | (0.34) |
| Price | **0.56\*\*\*** | **0.55\*\*\*** | **0.55\*\*\*** | **0.28\*\*\*** | **0.28\*\*\*** | **0.28\*\*\*** | 0.00 | 0.00 | 0.00 |
|  | **(31.37)** | **(31.65)** | **(31.66)** | **(12.59)** | **(12.77)** | **(12.77)** | (-0.12) | (-0.28) | (-0.28) |
| TxnVol | **0.77\*\*\*** | **0.78\*\*\*** | **0.78\*\*\*** | 0.05 | 0.05 | 0.05 | -0.02 | -0.02 | -0.02 |
|  | **(12.95)** | **(13.03)** | **(13.03)** | (0.62) | (0.68) | (0.68) | (-0.60) | (-0.58) | (-0.58) |
| BlockCnt | **1.51\*\*\*** | **1.51\*\*\*** | **1.51\*\*\*** | **2.58\*\*\*** | **2.57\*\*\*** | **2.57\*\*\*** | 0.05 | 0.04 | 0.04 |
|  | **(7.22)** | **(7.20)** | **(7.20)** | **(9.82)** | **(9.83)** | **(9.83)** | (0.34) | (0.30) | (0.30) |
| BlockTime | **1.33\*\*\*** | **1.33\*\*\*** | **1.33\*\*\*** | **2.70\*\*\*** | **2.69\*\*\*** | **2.69\*\*\*** | 0.11 | 0.10 | 0.10 |
|  | **(5.84)** | **(5.82)** | **(5.82)** | **(9.44)** | **(9.45)** | **(9.45)** | (0.71) | (0.67) | (0.67) |
| AvgFeeUsd | -0.03 | -0.04 | -0.04 | **-0.12\*** | **-0.12\*** | **-0.12\*** | 0.03 | 0.03 | 0.03 |
|  | (-0.58) | (-0.64) | (-0.64) | **(-1.67)** | **(-1.75)** | **(-1.74)** | (0.71) | (0.73) | (0.73) |
| N | 764 | 764 | 764 | 764 | 764 | 764 | 764 | 764 | 764 |
| Adj. R-sq | 0.75 | 0.75 | 0.75 | 0.33 | 0.33 | 0.33 | 0.01 | 0.00 | 0.00 |

Note: In this table, we run the regression model: $stock_{i,t} = \beta_0 + \beta_1 bribing_t + \beta_2 control_{i,t} + \beta_3 post_t \times bribing_t + \varepsilon_{i,t}$. In Columns (1) – (3), the dependent variable is price of the stock indice, and we use proxy benchmark, A and B, respectively. In Columns (4) – (6), the dependent variable is volume of the stock indice, and we use proxy benchmark, A and B, respectively. In Columns (7) – (9), the dependent variable is trading volume of the stock indice, and we use proxy benchmark, A and B, respectively. T-statistics are reported in parentheses. *, **, and *** denote significance levels at the 10%, 5%, and 1% levels based on the standard t-statistics.

**Table 12. Descriptive statistics of bribing proxies (after excluding low-value transactions)**

|  | Benchmark1 | Benchmark2 | A1 | A2 | B1 | B2 |
|---|---|---|---|---|---|---|
| **Mean** | 76.65 | 76.65 | 4858.84 | 4858.84 | 4803.37 | 4803.37 |
| **Median** | 4.57 | 4.57 | 17.66 | 17.66 | 10.86 | 10.86 |
| **Max** | 19708.69 | 19708.69 | 1361047.61 | 1361047.61 | 1361045.94 | 1361045.94 |
| **Min** | 0.04 | 0.04 | 0.10 | 0.10 | 0.00 | 0.00 |
| **Std** | 631.33 | 631.33 | 53103.51 | 53103.51 | 53103.10 | 53103.10 |

Note: This table reports the descriptive statistics of proxies of the active level of potential bribes. For transactions detected as potential bribes, we have two thresholds of value: 0.1 and 1. Then, bribing proxies will be calculated after deleting low-value transactions. For example, column 'benchmark1' is about proxy benchmark after deleting transaction with a value lower than 0.1.

**Appendices**

**Table A. 1. Definition of factors of cryptocurrencies**

| Factor abbreviation | Definition |
|---|---|
| **Token** | Token's price in USD |
| **Token.V** | Daily volume of the token |
| **Token.M** | The market capitalization of the token |
| **Token.R** | Daily return of the token |
| **V2 – V7** | 2-day – 7-day volatility of the token |

Note: In this paper, we focus on Ether (ETH), Bitcoin (BTC), Binance Coin (BNB), Binance USD (BUSD), Dai (DAI), Dogecoin (DOGE), Litecoin (LTC), Tether (USDT) and USD Coin (USDC).

**Table A. 2. Definition of transaction statistics**

| Factor abbreviation | Definition |
|---|---|
| **TxnVol** | Daily volume (in native units) of transactions |
| **TxnVolUsd** | Daily volume of transactions |
| **TxnCnt** | Daily number (in USD) of transactions |
| **TxnSize** | Total value (in native units) of transactions divided by the number of transactions |
| **TxnSizeUsd** | Total value (in USD) of transactions divided by the number of transactions |
| **TotalFee** | Total transaction fees (in native units) daily |
| **TotalFeeUsd** | Total transaction fees (in USD) daily |
| **AvgFee** | Average transaction fees (in native units) daily |
| **AvgFeeUsd** | Average transaction fees (in USD) daily |
| **BlockCnt** | The number of validated blocks daily |
| **BlockTime** | Average time (in seconds) between blocks per day |

**Table A. 3. Definition of network factors**

| Factor abbreviation | Definition |
|---|---|
| **Unique** | The number of unique addresses in blockchain |
| **New** | Daily new addresses in blockchain |
| **Active** | Daily active addresses, i.e., addresses that make a transaction, in blockchain |
| **Active.Ratio** | The percentage of addresses with a balance that make a transaction |

**Table A. 4. Factors of stock markets**

| Factor abbreviation | Definition |
|---|---|
| **S&P500** | Daily price of S&P500 |
| **S&P500.Vol** | Daily volume of S&P500 |
| **S&P500.R** | Daily return of S&P500 |
| **NASDAQ** | Daily price of NASDAQ |
| **NASDAQ.Vol** | Daily volume of NASDAQ |
| **NASDAQ.R** | Daily return of NASDAQ |
| **N225** | Daily price of Nikkei 225 (N225) |
| **N225.Vol** | Daily volume of N225 |
| **N225.R** | Daily return of N225 |
| **SSE** | Daily price of SSE Composite Index (SSE) |
| **SSE.Vol** | Daily volume of SSE |
| **SSE.R** | Daily return of SSE |